\documentclass[%
 reprint,
 amsmath,amssymb,
 aps,
]{revtex4-2}
\usepackage{color}
\usepackage{ulem}
\usepackage{graphicx}
\usepackage{dcolumn}
\usepackage{bm}

\usepackage{xparse,soul}
\makeatletter
\NewDocumentCommand{\sotwo}{O{red}O{black}+m}
    {%
        \begingroup
        \color{#1}%
        \setul{-.5ex}{.4pt}%
        \def\SOUL@uleverysyllable{%
            \rlap{%
                \color{#2}\the\SOUL@syllable
                \SOUL@setkern\SOUL@charkern}%
            \SOUL@ulunderline{%
                \phantom{\the\SOUL@syllable}}%
        }%
        \ul{#3}%
        \endgroup
    }
\makeatother

\begin{document}

\preprint{APS/123-QED}

\title{Domain Wall Conductivity with strong Coulomb interaction of two-dimensional massive Dirac Electrons in the Organic Conductor $\alpha$-(BEDT-TTF)$_2$I$_3$}

\author{D. Ohki$^1$}
\email{dohki@s.phys.nagoya-u.ac.jp}
\author{Y. Omori$^2$}%
\author{A. Kobayashi$^1$}
 \affiliation{$^1$Department of Physics, Nagoya University, Furo-cho, Chikusa-ku, Nagoya, 464-8602 Japan \\
  $^2$Toyota College, National Institute of Technology, Eisei-cho 2-1, Toyota, 471-8525 Japan \\
}%




\date{\today}

\newcommand{\ohki}[1]{\textcolor{black}{#1}}
\newcommand{\omori}[1]{\textcolor{black}{#1}}
\newcommand{\ohkimod}[1]{\textcolor{black}{#1}}
\newcommand{\ohkimodmod}[1]{\textcolor{black}{#1}}
\newcommand{\ohkiadd}[1]{\textcolor{black}{#1}}
%

\begin{abstract}
Motivated by the results of recent transport and optical conductivity
 studies, we propose a semi-infinite two-dimensional lattice model for
 interacting massive Dirac electrons in the pressurized organic
 conductor $\alpha$-(BEDT-TTF)$_2$I$_3$, and address the problem of
 domain wall conductivity in a charge-ordered insulating phase under
 realistic experimental conditions. Using the extended Hubbard model at
 a mean field level, we present results of extensive numerical studies
 around the critical region of the model, reporting on the resistivity
 and optical conductivity \omori{calculated} by means of the Nakano-Kubo
 formula. 
We find that the activation gap extracted from \ohkimodmod{the} resistivity data \ohkimodmod{can} be much smaller than the optical gap in the critical region, which is induced by metallic conduction along \ohkimodmod{an} one-dimensional domain wall emerging at the border of two charge-ordered ferroelectric regions with opposite polarizations.
 The data are consistent with the observed transport gap in real
 $\alpha$-(BEDT-TTF)$_2$I$_3$ samples that is reduced remarkably faster
 than the optical gap upon suppressing charge order with
 pressure. \ohkimodmod{Our optical conductivity also reveals} an additional shoulder-like structure \omori{at low energy inside the gap}, which is argued to be directly relevant to the metallic bound states residing on the domain wall.
\end{abstract}

\maketitle


\section{\label{sec:level1}Introduction}

%
 The quasi-two-dimensional (quasi-2D) electron system in the layered
 organic salt $\alpha$-(BEDT-TTF)$_2$I$_3$ has a unique
 pressure-temperature ($P$-$T$) phase diagram in which a 2D massless Dirac electron phase appears at high $P$~\cite{Kajita1992, Tajima2000, Kobayashi2004, Katayama2006, Kobayashi2007, Goerbig2008, Kajita2014} which an insulating phase is stabilized at low $P$ showing charge ordering. 
In the high-$P$ phase the space and time inversion symmetry guaranties the stability of the (spin-degenerate) Dirac points in the momentum space, and the 3/4-filling of the electronic band fixes the Fermi energy at the band-crossing \omori{points}. 
In contrast, the inversion symmetry is spontaneously broken in the low-$P$ insulating phase where electrons are localized and form a \omori{stripe-type} charge-ordering pattern along the crystalline \omori{$b$} axis~\cite{KinoFukuyama, Seo2000, TakahashiStripe, Kakiuchi2007}. 
At ambient pressure the charge\ohkimodmod{-}ordered phase appears below a transition
 temperature \ohkimodmod{of} $T_{\rm CO}=135$ K. \omori{An} application of a hydrostatic
 pressure linearly reduces $T_{\rm CO}$ and eventually suppresses the
 phase transition above a critical pressure $P_{\rm c} \ohkimodmod{\simeq} 12$ kbar,
 stabilizing the massless Dirac electron phase at low temperature. 

%
A narrow energy bandwidth characteristic for this type of organic
conductors gives rise to strong electronic correlation effects in both phases \cite{Kajita1992, Tajima2000, Bender1984}. 
As theoretically predicted and also experimentally confirmed, the electron-electron Coulomb interaction plays a significant role in the stripe-type charge\ohkimodmod{-}ordered phase \ohkimodmod{at low-$P$}. \omori{For} example, recent NMR and Monte Carlo studies \ohkimodmod{point to} a spin-excitation nature that is consistent with \ohkimodmod{one-dimensional (1D)} alternating Heisenberg spin chains~\cite{KinoFukuyama, Seo2000,Tanaka2016,Ishikawa2016}. 
\ohkimodmod{Moreover, a novel charge-ordered phase accompanied by massive Dirac electrons has been predicted in the vicinity of the critical region of the phase diagram ($P \simeq P_c$), which is induced by the short-range part of the Coulomb interaction (This charge-ordered massive Dirac electron phase can be distinguished from the ordinary charge-ordered phase at lower $P$ in terms of the valley Hall effect since the former has a finite valley Chern number, whereas the latter has none~\cite{Katayama2006, Kobayashi2007, Kobayashi2008IOP, Dietl, Montambaux2009PRB, Montambaux2009EPJB, Kobayashi2011, Matsuno2016, Omori2017,Ohki2018JPSJ, Ohki2018Crystals}).} 
In the \ohkimodmod{high-$P$} massless Dirac electron phase, not only the short-range repulsive interactions but also the long-range part of the Coulomb interaction (appearing due to the absence of metallic screening at the Dirac point) induce various anomalies in \ohkimodmod{the} NMR spin susceptibilities: \omori{A ferrimagnetic spin polarization, a logarithmic suppression of the Knight shift, and \ohkimodmod{orders of magnitude} enhancement of the Korringa ratio~\ohkimodmod{\cite{Hirata2016, Hirata2017}}. Moreover,} at low temperature signatures of inter-valley excitonic spin fluctuations were reported as a precursor to the \ohkiadd{excitonic transition \ohkimodmod{by the long-range part of the Coulomb interaction}~\cite{Hirata2016, Matsuno2017, Hirata2017, Matsuno2018}}.

 Recently, a surprising transport property under pressure has been reported by resistivity measurements~\cite{Liu2016}\ohkimodmod{. The} transport gap \omori{$\Delta_\rho$ estimated} by Arrhenius plot\ohkimodmod{s turns out to have a} much smaller \ohkimodmod{value compared to} \ohkimodmod{the} \ohkimodmod{optical} gap \omori{$\Delta_{\rm O}$} \ohkimodmod{extracted from} optical conductivity measurements~\cite{Beyer2016}. 
\omori{The resistivity gap} $\Delta_{\rm \rho}$ is strongly suppressed as pressure is increased and becomes zero at $P \cong 7$ kbar \omori{while} $T_{\rm CO}$ and \omori{the optical gap} $\Delta_{\rm O}$ remain finite until the pressure reaches $P_c$. 
As a candidate of \ohkimodmod{a} possible dc conduction mechanism \omori{in this critical region \ohkimodmod{(}$7$ kbar $\lesssim P \lesssim P_c$\ohkimodmod{)}}, \ohkimodmod{a} 1D conduction \ohkimodmod{scheme} has been predicted along a gapless bound state on a domain wall \ohkimodmod{formed} between two \omori{kinds of} charge-ordered domains \omori{with different polarization}~\cite{Matsuno2016, Omori2017,Ohki2018JPSJ}. 
\ohkimodmod{Interestingly, \ohkimodmod{the charge ordering} in $\alpha$-(BEDT-TTF)$_2$I$_3$ is also found to be accompanied by ferroelectricity~\cite{Yamamoto2008,Yamamoto2010,Lunkenheimer} as well as $180^\circ$ polar domains having a domain size of several hundred micrometers~\cite{Yamamoto2010}.} 
However, \ohkimodmod{it remains unclear how the} formation \ohkimodmod{of these domain walls is related to} the observed contrasting pressure dependence of $\Delta_{\rm \rho}$ and $\Delta_{\rm O}$. 
\ohkimodmod{To construct a realistic theory in this critical region, one is therefore motivated to start with a minimal model that has a single domain wall formed between two ferroelectric domains possessing opposite electric polarization.}

 In this paper we develop a numerical approach which accounts for the distinct pressure dependence of $\Delta_{\rm \rho}$ and $\Delta_{\rm O}$ in the pressurized $\alpha$-(BEDT-TTF)$_2$I$_3$, using a cylindrical boundary condition that naturally introduces a domain wall in a space-dependent mean-field theory~\cite{Hasegawa2011, Omori2014, Matsuno2016, Omori2017,Ohki2018JPSJ}. 
The interaction between electrons is treated within the extended Hubbard model, where the canonical on-site interaction and the nearest-neighbor interactions are included. 
In our recent mean-field studies \omori{with} semi-infinite~\cite{Matsuno2016, Omori2017,Ohki2018JPSJ} and 2D periodic~\cite{Ohki2018Crystals} boundary conditions, \ohkimodmod{indications were found that} the influence of pressure \ohkimodmod{can} be parametrized by the strength of the nearest-neighbor Coulomb interaction along the crystalline \omori{$a$} axis $V_a$, which varies most sensitively upon changing the applied external pressure~\cite{Kobayashi2009IOP, Tajima2009} and plays dominant roles in stabilizing \ohkimodmod{the} stripe-type charge order~\cite{Seo2000}. 
Following this hypothesis, we utilize $V_a$ as our control parameter \ohkimodmod{in this study}. 
Periodic boundary conditions \ohkimodmod{are considered} in the \omori{$a$} direction but edges \ohkimodmod{are placed} in the \omori{$b$} direction \omori{to} \ohkimodmod{introduce a} \omori{domain wall in} \ohkimodmod{the} \omori{model.} 
We present extensive numerical calculations of the dc resistivity and the optical conductivity using the T-matrix approximation \ohkimodmod{combined with} the Nakano-Kubo formula~\cite{Streda, Shon, Proskurin, Ruegg2008, Omori2017,Ohki2018Crystals}.  \ohkimodmod{This approach provides us} a novel way to understand the anomalous \omori{behavior} of \ohkimodmod{the experimentally reported transport and optical} \omori{gaps} ($\Delta_{\rm \rho} \omori{\ll} \Delta_{\rm O}$ around $P \sim 7$ kbar~\cite{Liu2016, Beyer2016}) \ohkimodmod{by} domain wall conduction along the \omori{$a$} direction in the model's critical region. 
We also find an unexpected shoulder-like structure in the optical conductivity at low energy, which is discussed in terms of a \ohkimodmod{1D} metallic bound state on the domain wall.

The remainder of \ohkimodmod{this} paper is organized as follows. 
\omori{In Sec.~II.~A,} we lay out the extended Hubbard model for $\alpha$-(BEDT-TTF)$_2$I$_3$ \ohkimodmod{subjected to} two kinds of cylindrical boundary conditions either symmetric or asymmetric in the \omori{$b$} direction~\cite{Matsuno2016, Omori2017,Ohki2018JPSJ}. 
We also present \ohkimodmod{a} summary of \ohkimodmod{earlier} works in this section, \ohkimodmod{putting particular emphasis} on \ohkimodmod{geometrical mechanisms behind the formation} of \ohkimodmod{a single} \omori{domain wall and \ohkimodmod{showing} its energy spectrum. In Sec.~II.~B, we} present the formalisms of dc and optical conductivities within the T-matrix approximation using the Nakano-Kubo formula. 
\ohki{Details of \omori{their} \ohkimodmod{formulations} are summarized in Appendix A.} 
Our numerical results are shown in Sec.~III. \omori{In Sec.~III.~A, we} \ohki{first focus on} \ohkimodmod{the} \omori{temperature dependence} \ohki{of} \ohkimodmod{the spatially}\ohki{-resolved electronic states} \ohki{for the two types of boundary conditions}
\omori{with} \ohkimodmod{and} \omori{without \ohkimodmod{a} domain wall. We also present} \ohki{the interaction-temperature} \ohkimodmod{($V_a$-$T$)} \omori{phase diagram which}\ohkimodmod{, for the symmetric-edge case, confirms the presence of a domain wall in a wide parameter region in the charge-ordered phase}. 
\omori{In Sec.~III.~B,} \ohki{results} \ohkimodmod{for} \ohki{the temperature} \ohkimodmod{dependent} \ohki{dc resistivity} \ohkimodmod{are shown} \ohki{for various sizes of} \ohkimodmod{$V_a$, in which Arrhenius fits to the data reveal a} transport gap $\Delta_{\rm \rho}$ \ohkimodmod{that is strongly dependent on the fitted temperature range}. 
\omori{In Sec.~III.~C, we} evaluate the optical conductivity as a function of energy and boundary types, 
which provides the optical gap $2\Delta_{\rm O}$ for a range of interaction sizes. 
We summarize our findings for the domain wall conduction in Sec.~VI and also try to associate our proposed conduction mechanism to the results in real $\alpha$-(BEDT-TTF)$_2$I$_3$ by considering realistic experimental and materialistic conditions.

\section{Model and Formulation}
\vspace{-1cm}
\ohkimodmod{
\subsection{Models and Summary of Previous Studies: Emergent Domain Wall by Geometrical Constraints}
}

\begin{figure}
\begin{centering}
\includegraphics[width=75mm]{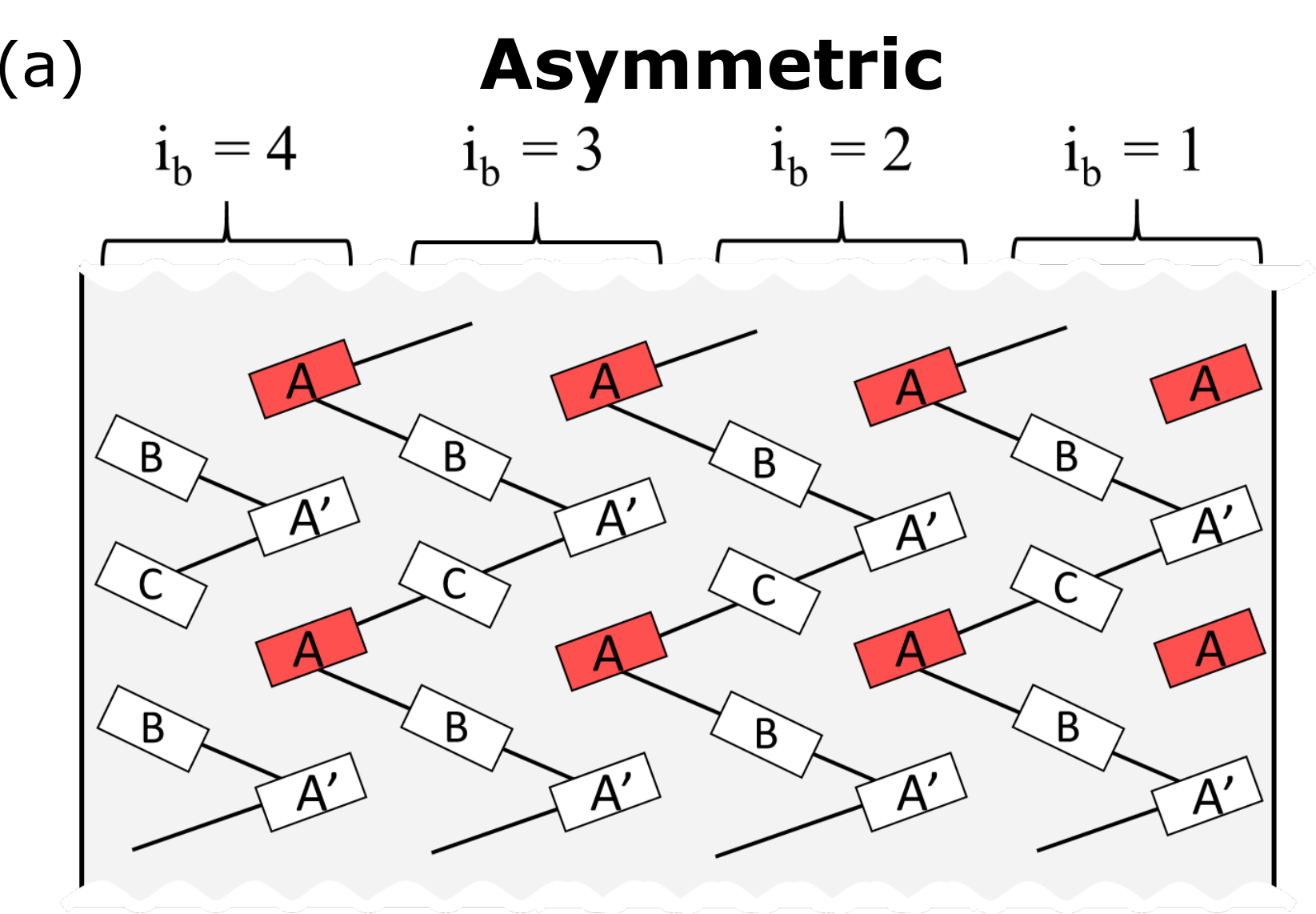}
\includegraphics[width=75mm]{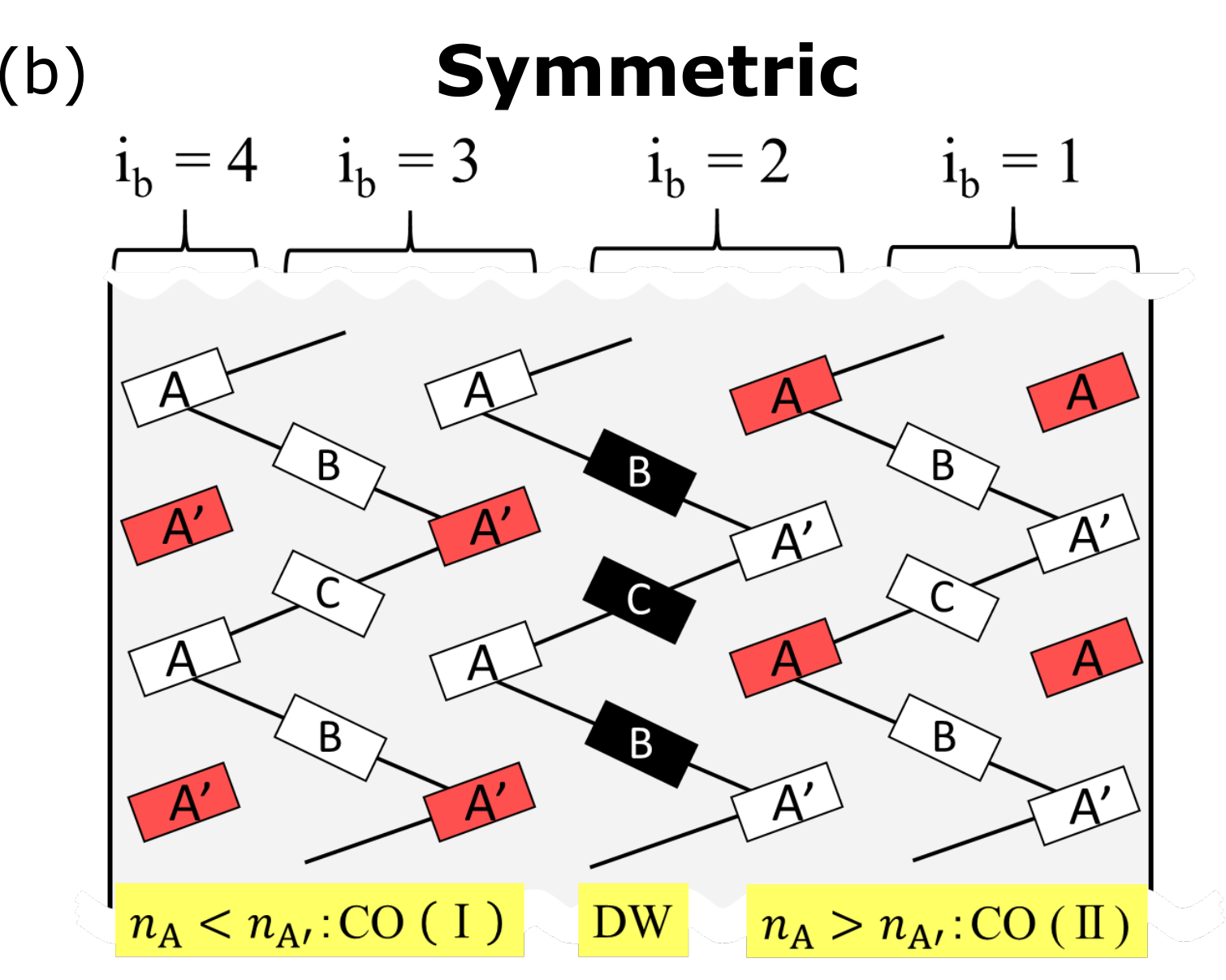}
\includegraphics[width=80mm]{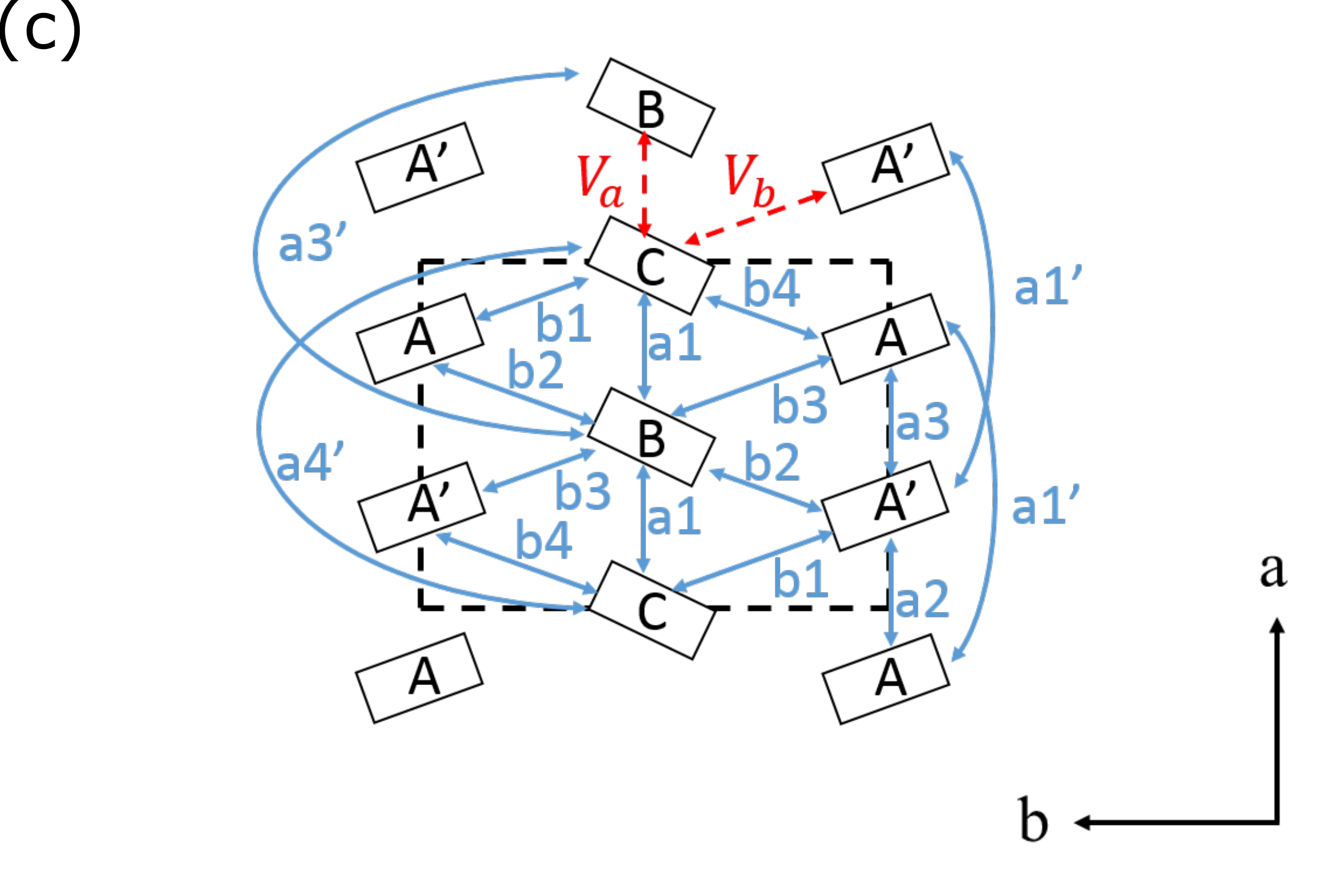}
\caption{\label{illustration}(Color online) 
Semi-infinite boundary condition in a model system of $\alpha$-(BEDT-TTF)$_2$I$_3$, which is periodic in the \omori{$a$} direction and has edges in the \omori{$b$} direction. 
The rectangles indicate A, A$'$, B, and C sites corresponding to the nonequivalent BEDT-TTF molecules in the 2D unit cell (in the charge-ordered phase). 
Two edge structures are considered in this study; (a) an asymmetric-edge pattern having either AA$'$ or BC columns at each edge \ohkimodmod{(AA$'$-BC)} and (b) a symmetric-edge pattern having AA$'$ columns on both edges \ohkimodmod{(AA$'$-AA$'$) (For the sake of simplicity, only a minimum number of AA$'$ columns needed to introduce a domain wall is illustrated)}. 
The black bold lines stand for the network of largest hopping integrals forming a zigzag chain along the $a$ axis. 
The red filled rectangles indicate the \omori{A or A$'$} molecules where the mean electronic density becomes large in the charge-ordered phase, while the black ones represent \omori{B and C molecules} that constitute a domain wall~\cite{Omori2017}, \omori{as shown in Fig.~2(d)}. 
(c) The 2D network of nearest-neighbor and next-nearest neighbor transfer integrals in the conducting phase (solid arrows), $t_{a1}, t_{a2},\cdots , t_{a4'}, t_{b1},\cdots , t_{b4}$. 
The nearest-neighbor Coulomb repulsions (dashed arrows) $V_a$ and $V_b$ are also indicated. 
}
\end{centering}
\end{figure}

%
\ohki{
Before presenting} \ohkimodmod{the main results} \ohki{of this} \ohkimodmod{paper}, \ohkimodmod{let us first} \ohki{introduce our} \ohkimodmod{semi-infinite models that require special cares on the treatment of edges. We then proceed to show the Hamiltonian} \ohki{and} \ohkimodmod{summarize} \ohki{previous} \ohkimodmod{key findings} \ohki{derived from} \ohkimodmod{these models; in particular, we focus} \ohkimodmod{on the geometry-necessitated} \ohki{mechanism of domain wall} \ohkimodmod{formation} \ohki{between two different charge-\ohkimodmod{ordered} domains~\cite{Matsuno2016, Omori2017,Ohki2018JPSJ}.} 

\ohkimodmod{To begin with, the model we rely on aims to describe conduction} mechanisms in $\alpha$-(BEDT-TTF)$_2$I$_3$ in a realistic \ohkimodmod{situation} where cylindrical boundary conditions are employed in the conducting 2D plane. \ohkimodmod{More specifically, we} assume a periodic \omori{boundary} condition along the crystalline \omori{$a$} axis, whereas edges are introduced along the \omori{$b$} axis, as presented in \omori{Figs.}~1(a) and (b). 
The 2D unit cell in the \ohkimodmod{(low-$P$)} charge-ordered state contains four nonequivalent molecular sites \ohkimodmod{(}dubbed A, A$'$, B and C\ohkimodmod{)}~\cite{Bender1984, Kondo2009} which form two distinct columns in the \omori{$a$} direction, \ohkimodmod{labeled as AA$'$ and BC (see Fig.~1)}. 
Because of these two column types, there are three ways to introduce edges at the two ends in the $b$ direction: AA$'$-AA$'$, AA$'$-BC and BC-BC. 
\ohkimodmod{Among these,} \ohki{AA$'$-AA$'$ and BC-BC \ohkimodmod{have} symmetric edges} \ohkimodmod{and lead to} \ohki{similar results} \ohkimodmod{within this theory.} 
\omori{Therefore}\ohkimodmod{, we will not distinguish them hereafter and will only focus on the two edge types of} AA$'$-BC and AA$'$-AA$'$ \ohkimodmod{that are} either asymmetric (AA$'$-BC) \ohkimodmod{[Fig.~1(a)]} or symmetric (AA$'$-AA$'$) \ohkimodmod{[Fig.~1(b)]} in the \omori{$b$} direction. 

The electronic structure \ohkimodmod{in} \omori{$\alpha$-(BEDT-TTF)$_2$I$_3$ is} \ohkimodmod{somewhat} involved \ohkimodmod{due to} a complicated \omori{2D} inter-molecular network of hopping integrals\ohkimodmod{. Figure 1(c) shows the} nearest and next-nearest neighbor hoppings \ohkimodmod{in} the conducting \ohkimodmod{phase used in this study}, where the system has inversion centers between the molecules A and A$'$ as well as on the molecules B and C~\cite{Kakiuchi2007}. 
\ohkimodmod{Note} \omori{that the} sizeable hopping integrals \ohkimodmod{[in particular} b1 and b2 in Fig.~1(c)\ohkimodmod{]} \ohkimodmod{form} a zigzag network along the \omori{$a$} direction \ohkimodmod{[shown by} solid lines in Figs.~1(a) and 1(b)\ohkimodmod{]}. 

\vspace{-0.41cm}
\ohkimodmod{As we reported previously,} the molecules at the edges residing off these zigzags are close to charge neutral (i.e., closed shell), whereas the other molecules residing on the zigzags are positively charged reflecting the hole 1/4 filling (electron 3/4 filling) of the electronic band\ohkimodmod{s}~\cite{Omori2017}. 
That kind of charge neutral molecules isolated from the zigzags appear only in the AA$'$ column locating at the edges; for the \ohkiadd{(AA$'$-AA$'$)} symmetric-edge pattern\ohkimodmod{, the isolation occurs at} the molecule A \omori{on one edge} \ohkimodmod{while it takes place at} the molecule A$'$ on the other \ohkimodmod{[Fig.~1(b)]}. 
Because of this and the fact that the charge carriers (i.e., holes) are localized either on the molecule A or A$'$ \omori{(plus B)} in the strip-type charge-ordered state~\cite{Kakiuchi2007}, the \ohkiadd{(AA$'$-AA$'$)} symmetric-edge pattern inevitably acquires at least one domain wall on a BC column \ohkimodmod{necessitated by the geometry}~\omori{\cite{Omori2017}} \ohkimodmod{[}\omori{Figs.~1(b) and 2(d)}\ohkimodmod{]}. 
For the \ohkiadd{(AA$'$-BC)} asymmetric-edge pattern, by contrast, no domain walls are
required as there is only one AA$'$ column at \ohkimodmod{an} edge, and the charge-ordering pattern can be uniquely determined \ohkimodmod{[}Fig.~1(a)\ohkimodmod{]}. 
\ohkimodmod{Generally speaking,} charge-ordered domain\ohkimodmod{s are} accompanied by spontaneous electric polarization~\cite{Yamamoto2008,Yamamoto2010}, \ohkimodmod{causing} multiple domain walls at the boundaries of \ohkimodmod{several} ferroelectric domains, as is the case for conventional ferroelectric materials. 
\ohkimodmod{To make the story simple, however,} we will omit these \ohkimodmod{additional} domain walls in this study and concentrate only on the one required from the \ohkimodmod{geometrical} arguments. 

\vspace{-0.41cm}
\ohkimodmod{Next, so as to the Hamiltonian, we consider} a 2D Hubbard-type model \ohkimodmod{serving} as a standard framework to study interacting electrons on a lattice. 
In the orthodox Hubbard model one only considers a repulsive on-site interaction between electrons of opposite spin, but in this study we incorporate the nearest-neighbor interactions playing one of the most essential roles in driving the charge-ordering transition in
$\alpha$-(BEDT-TTF)$_2$I$_3$, as \omori{theoretically suggested~\cite{Seo2000} and \ohkimodmod{experimentally} confirmed by NMR~\cite{Ishikawa2016}.} 
We also introduce edge potentials following Ref.~\omori{\cite{Omori2017}} to treat the effects of surface charge recombination. (In addition, we note that the interactions between BEDT-TTF molecules and I$_3$ anions would also play some roles in charge order as argued in Ref.~\cite{Alemany}. Although we do not exclude that possibility, these interactions are assumed to have only minor \ohkimodmod{influences} on our arguments and are thus neglected.) 
The Hubbard-type model used in this study is described by the Hamiltonian $H=H_{\rm kin}+H_{\rm int}$ which is given by a sum of the kinetic term 
\begin{eqnarray}
H_{\rm kin}
= {\sum_{\langle l, l' \rangle}} {\sum_{\sigma \sigma'}} \left( t_{l, l'} a^\dag_{l \sigma} a_{l' \sigma'} + {\rm h.c.} \right) ,
\end{eqnarray}
and the interaction term 
\begin{eqnarray}
{H_{\rm int}}&=&{\sum_{l}} U n_{l \uparrow}n_{l \downarrow} + {\sum_{\langle l, l' \rangle}} {\sum_{\sigma \sigma'}} V_{l, l'} n_{l \sigma}n_{l' \sigma'} \nonumber\\
&&+ {\sum_{edge}} V_{\rm edge} n_{l}.
\end{eqnarray}
Here, $ l = (i_a, i_b, \alpha)$ represents a molecule $\alpha$ = A,
A$'$, B, and C in the $(i_a, i_b)$-th unit cell, for the space slices in
the \omori{$a$} direction $i_a = 1, \cdots , N_a$ and in the \omori{$b$}
direction $i_b = 1, \cdots , N_b$\ohkimodmod{, respectively}. $t_{l, l'}$ is the nearest\ohkimodmod{-neighbor} or next-nearest-neighbor hopping integral between sites $l$ and $l'$. $U$ is the
(molecule-independent) on-site interaction, \omori{and} $V_{l, l'}$ \ohkimodmod{represents} the
nearest-neighbor \ohkimodmod{repulsive} interaction between electrons at sites $l$ and $l'$, for which \ohkimodmod{one can consider} two types\ohkimodmod{, dubbed} \omori{$V_a$ and $V_b$ as shown in Fig.~1(c).} 
${a_{l \sigma}}^\dag$ (${a_{l\sigma}}$) creates (annihilates) an electron of spin $\sigma = \uparrow
, \downarrow$ at site $l$, $n_{l \sigma} = {a_{l \sigma}}^\dag a_{l
\sigma}$ stands for the density of electrons with spin $\sigma$ at site
$l$, and the density $n_l = \sum_\sigma n_{l \sigma}$ at site $l$ \omori{is} summed over all spin projections. 
Following recent analyses of the high-pressure NMR data~\cite{Hirata2016, Hirata2017}, we use the values of $t_{l, l'}$ given by a first principle calculation in Ref.~\cite{Kino}: $t_{a1} = -0.0267$, $t_{a2} = -0.0511$, $t_{a3} = 0.0323$, $t_{b1} = 0.1241$, $t_{b2} = 0.1296$, $t_{b3} = 0.0513$, $t_{b4} = 0.0152$, $t_{a1}' = 0.0119$, $t_{a3}' = 0.0046$, $t_{a4}' = 0.0060$ \ohkimodmod{[}in eV; see Fig.~1(c)\ohkimodmod{]}. 
The interaction parameters are set to literature-accepted values $U =
0.4$~eV and $V_b = 0.05$~eV~\cite{Matsuno2016, Omori2017,Ohki2018JPSJ}, and\ohkimodmod{, again,} $V_a$ is the control parameter of the model. 
The third \omori{term} \ohkimodmod{in} $H_{\rm int}$ \ohkimodmod{[Eq. (2)]} \ohkimodmod{represents an edge potential $V_{\rm edge}$ that only} acts on the electrons \ohkimodmod{residing at the} edge sites\ohkimodmod{; following Ref.~\cite{Omori2017},} we \ohkimodmod{assume} $V_{\rm edge} = 4V_b$. 
The system size is set as $N_a = 200$ and $N_b = 60$\ohkimodmod{, and} a Fourier transformation \ohkimodmod{is performed} along the $a$ axis \ohkimodmod{using} the operator ${a_{l \sigma}}={N_a^{-1/2}{\sum_{k_a}}a_{k_a l_b \sigma}e^{ik_a i_a}}$ defined for $l_b = (i_b, \alpha)$ and the wavenumber vector $k_a$\ohkimodmod{. Within} the Hartree approximation we get \ohkimodmod{a mean-field} Hamiltonian $H_{\rm MF}$ \ohkimodmod{that is} diagonalized \ohkimodmod{in the} band representation and \ohkimodmod{has an} energy eigenvalue $E_{\nu \sigma}(k_a)$\ohkimodmod{, as described by} 
\begin{eqnarray}
H_{\rm MF}= \sum_{k_a \nu \sigma} E_{\nu \sigma}(k_a) c^\dag_{k_a \nu \sigma}c_{k_a \nu \sigma}+{\rm const.} ,
\\
E_{\nu \sigma}(k_a)d_{l_b \nu \sigma}(k_a) = \sum_{{l_b}'} \tilde{\varepsilon}_{l_b {l_b}' \sigma}(k_a)d_{{l_b}' \nu \sigma}(k_a) ,
\end{eqnarray}
\ohkimodmod{where} $\nu$ indicates \ohkimodmod{the} band index and $c_{k_a \nu \sigma}$ is defined using
an unitary matrix $d_{l_b \nu \sigma}(k_a)$ \ohkimodmod{[which is the eigenfunction in Eq. (4)]} as $c_{{k_a} \nu \sigma}=\sum_{l_b'} d^*_{l_b' \nu \sigma}({k_a}) a_{{k_a} l_b' \sigma}$. 
The \ohkimodmod{charge carrier density} at \ohkimodmod{the} site $l_b$ is calculated as $\langle n_{l_b} \rangle = N_a^{-1} \sum_{k_a \sigma} \sum_{\nu}
\omori{ \left| d_{l_b \nu \sigma}({k_a}) \right|^2 }
\langle c^\dag_{{k_a} \nu \sigma}c_{{k_a} \nu \sigma} \rangle$.
\ohkimodmod{Throughout this manuscript, } \omori{eV} \ohkimodmod{is used} \omori{as the unit of energy.}

\begin{figure}
\begin{centering}
\includegraphics[width=88mm]{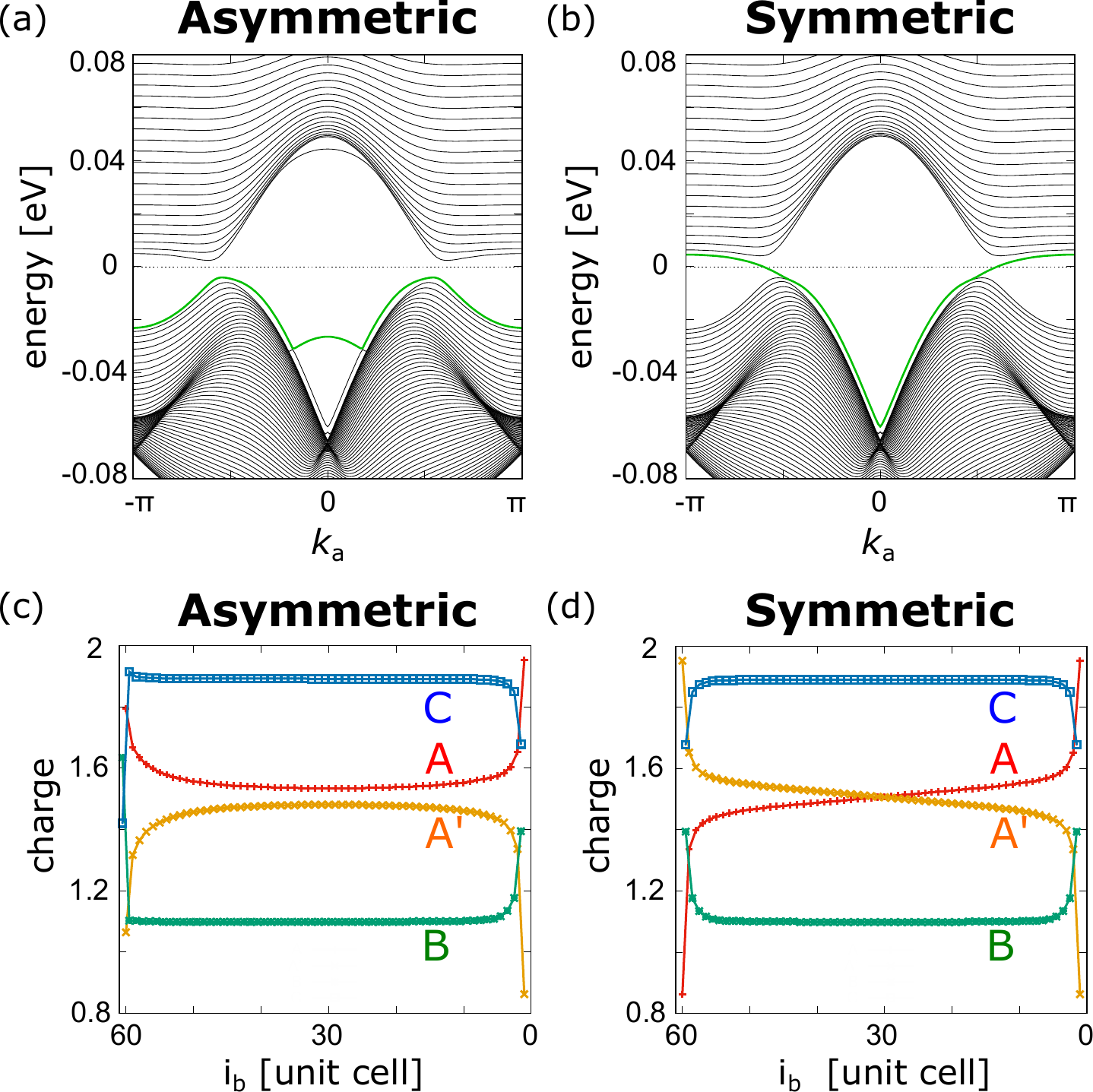}
\caption{\label{illustration}(Color online) 
The energy spectrum near the Fermi energy at $V_a = 0.21$ and $T = T^* =
 0.0075$ plotted as a function of the wavevector in the \omori{$a$} direction ($k_a$) for the \ohkimodmod{(AA$'$-BC)} asymmetric-edge pattern (a) and the \ohkimodmod{(AA$'$-AA$'$)} symmetric-edge pattern (b). 
The corresponding mean electron charge density at each molecule in the unit cell $i_b$ is also presented in (c) and (d), respectively. 
The green solid curves in (a) and (b) specify those bands that are
 important for the arguments of domain wall, labeled as $\nu = 180$ for
 the \ohkiadd{(AA$'$-BC)} asymmetric-\omori{edge} \ohkimodmod{(a)} and $\nu = 179$ for the \ohkiadd{(AA$'$-AA$'$)} symmetric-\omori{edge} \ohkimodmod{(b)}. 
}
\end{centering}
\end{figure}

\ohkimodmod{Finally,} \omori{let us show} \ohkimodmod{that an emergent} \omori{gapless} \ohkimodmod{state naturally appears on the the domain wall in the above model~\cite{Matsuno2016, Omori2017,Ohki2018JPSJ}.}
 Figures 2(a) and (b) show the mean-field energy eigenvalue $E_{\nu}(k_a)$ around the Fermi level at the \ohkimodmod{transition} temperature \ohkimodmod{to the charge-ordered state (hereafter referred to as $T^*$)} for the \ohkiadd{(AA$'$-BC)} asymmetric- and \ohkiadd{(AA$'$-AA$'$)} symmetric-edge patterns, respectively. 
\ohkimodmod{(Note that $T^*$ is defined from the point where a discontinuous jump occurs in the second derivative of the Helmholtz Free energy $\partial^2 F(T) / \partial T^2$.)} 
\ohkimodmod{A} gapless \omori{energy spectrum} arising from the \omori{domain-wall bound state} \ohkimodmod{crosses} $E_{\rm F}$ for the \ohkimodmod{(AA$'$-AA$'$)} symmetric-\omori{edge} [green curve in Fig.~2(b)], whereas \ohkimodmod{there is no gapless states} for the \ohkimodmod{(AA$'$-BC)} asymmetric-\omori{edge} \ohkimodmod{such that} the system \ohkimodmod{remains} fully gapped [\ohkimodmod{see} Fig.~2(a), where the green curve represents the top level in the valence band]. 
In \ohkimodmod{Figs.~2(c) and (d)} the corresponding mean electron density ${\langle}n_{l_b}{\rangle}$ at site $l_b = (i_b, \alpha)$ is plotted for the \ohkiadd{(AA$'$-BC)} asymmetric-\omori{edge [Fig.~2(c)] and} \ohkiadd{(AA$'$-AA$'$)} symmetric-\omori{edge [Fig.~2(d)] \ohkimodmod{patterns}}. 
\ohkimodmod{For} the \ohkiadd{(AA$'$-AA$'$)} symmetric \ohkimodmod{case} \ohkimodmod{$\langle n_{(i_b, A)}\rangle$ intersects $\langle n_{(i_b, A')}\rangle$} at $i_b = 30$ \ohkimodmod{signifying} the appearance of \ohkimodmod{a} domain wall\ohkimodmod{.} 
\ohkimod{In this study we} \ohkimodmod{investigate} \ohki{the temperature dependence of these electronic states}\ohkimodmod{, whose detailed nature including the phase diagram has not yet been fully understood}. 
\\
\\
\\

\omori{
\subsection{Optical Conductivity}
}

\ohkimod{In order to evaluate the optical gap, we} \ohkimodmod{calculate} \ohkimod{the optical conductivity given by} \ohkimodmod{the} \ohkimod{following Nakano-Kubo formula}
{
\begin{eqnarray}
\sigma^a(\omega)=\frac{1}{i\omega}\left[ Q^R(\omega) - Q^R(0) \right] ,
\end{eqnarray}
where $Q^R(\omega)$ is a current-current correlation function \ohkimodmod{in which} the Matsubara frequency $i\epsilon_n$ \ohkimodmod{is analytically connected} to a real frequency $\omega$. 
\ohkimodmod{The} optical conductivity $\sigma^a(\omega)$ \ohkimodmod{is calculated} in a clean limit. 
In the \ohkimodmod{zero-frequency} limit equation (5) gives the longitudinal conductivity \ohkimodmod{along the $a$ axis (corresponding to the Drude term)}~\cite{Streda, Shon, Proskurin, Ruegg2008, Omori2017} 
\begin{eqnarray}
\sigma^a = \int d\omega \left( -\frac{d f}{d\omega} \right)\Phi(\omega) ,
\end{eqnarray}
where $f$ is the Fermi-Dirac function and $\Phi(\omega)$ is \ohkimodmod{a distribution function that is calculated for impurity scatterings using the} T-matrix approximation, \ohkimodmod{as expressed by} 
\begin{equation}
\Phi(\omega)=\frac{4e^2}{\Omega}\sum_{k_a \nu}\left| {\rm v}^a_{\nu}(k_a) \right|^2 \tau_\nu (\omega, k_a)\delta(\hbar \omega - E_\nu (k_a)) .
\end{equation}
Here, \ohkimodmod{$\Omega = N_a \times N_b$ is the 2D system size,} $\tau_\nu (\omega, k_a)$ is an relaxation time 
and ${\rm v}^a_{\nu}(k_a)$ is a velocity \ohkimodmod{derived from a wavenumber derivative of} the energy eigenvalue \ohkimodmod{$E_{\nu}(k_a)$}. A spatially resolved conductivity $\sigma^a_{i_b}$ is calculated as
\begin{eqnarray}
\sigma^a_{i_b} &=& \int d\omega\left( -\frac{df}{d\omega} \right)\Phi_{i_b}(\omega) ,\\
\Phi_{i_b}(\omega) &=& \sum_{\alpha {l_b}'}\Phi_{l_b {l_b}'}(\omega) ,\nonumber\\
&=& \frac{4e^2}{N_a}\sum_{\alpha {l_b}'}\sum_{k_a \nu}{\rm v}^a_\nu (k_a)\left[ d^*_{l_b \nu}(k_a)v^a_{l_b {l_b}'}(k_a)d_{{l_b}' \nu}(k_a) \right]\nonumber\\
&&\times \hspace{0.2cm} \tau_\nu (\omega, k_a)\delta(\hbar \omega - E_\nu (k_a)) .
\end{eqnarray}
\ohkimodmod{(For the details of} derivation\ohkimodmod{s, see} Appendix A and \ohkimodmod{Ref~\cite{Omori2017}.)} In the following, the conductivity is \ohkimodmod{normalized to} the universal conductivity $\sigma_0 = 4e^2 /\pi h$\ohkimodmod{, and the Drude term is subtracted}.} 
\\
\\
\\

\section{Numerical Results}
\vspace{-1cm}
\omori{
\subsection{Local Electronic Structures and the Massive Dirac Electron Phase}
}
%

\begin{figure}
\begin{flushleft}
\includegraphics[width=88mm]{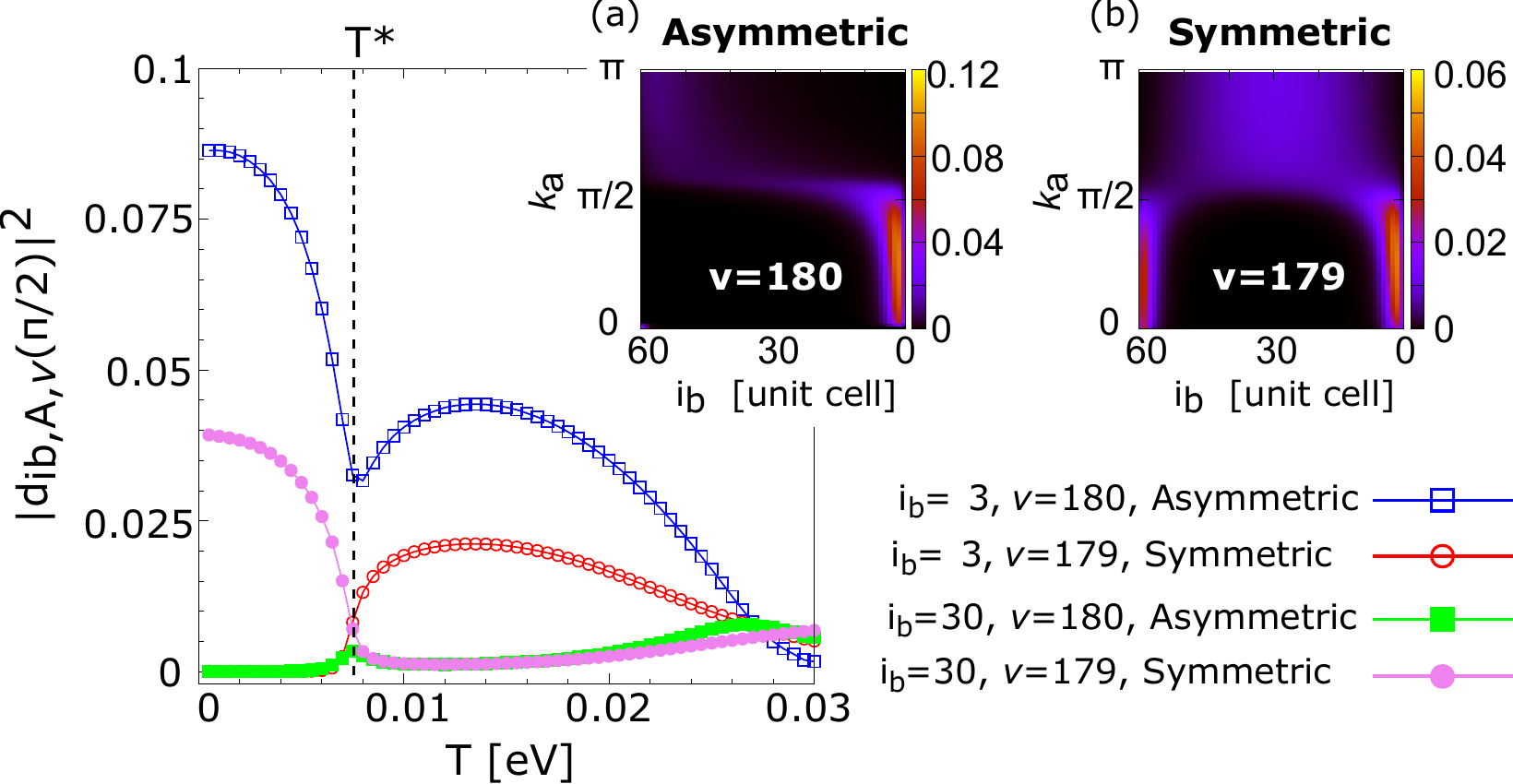}
\caption{\label{illustration}(Color online) 
Temperature dependence of the squared A-site \ohkimodmod{eigenfunction} at $k_a = \pi / 2$ and $V_a = 0.21$ plotted for the \ohkimodmod{(AA$'$-BC)} \omori{asymmetric}\ohkimodmod{-edge} ($\nu = 180$) and \ohkimodmod{(AA$'$-AA$'$)} \omori{symmetric}\ohkimodmod{-edge} ($\nu = 179$) \ohkimodmod{cases}. 
The results around the right edge ($i_b = 3$) and the center ($i_b = 30$) are presented. 
$T^*$ represents the charge-ordering transition temperature. 
Inset: the 2D plot of the squared A-site \ohkimodmod{eigenfunction} at $V_a = 0.21$ and
 $T = 0.011 (> T^*)$, plotted as a function of $k_a$ and $i_b$ for
 \omori{the} \ohkimodmod{(AA$'$-BC)} asymmetric ($\nu = 180$) (a) and \ohkimodmod{(AA$'$-AA$'$)} symmetric ($\nu = 179$) (b) edges. 
}
\end{flushleft}
\end{figure}
 In Fig.~3 we plot the temperature dependence of the modulus of \ohkimodmod{the} squared
 A-site \ohkimodmod{eigenfunction} at $k_a=\pi / 2$, $\left|d_{i_b A \nu}(\pi / 2)\right|^2$, \ohkimodmod{in which we focus} on the band\ohkimodmod{s} $\nu = 179$ and \ohkimodmod{$\nu = 180$ at} the unit cell \ohkimodmod{locating} at the center ($i_b = 30$) and around the right end ($i_b = 3$). \ohkimodmod{(For the definition of these band indices, see the caption of Fig.~2}). 
The 2D plot of $\left|d_{i_b A \nu}(\pi / 2)\right|^2$ at $T = 0.011 (> T^*)$ is also shown in the insets, plotted as a function of $i_b$ and $k_a$ for \omori{the} \ohkimodmod{(AA$'$-BC)} asymmetric-\ohkimodmod{edge pattern} [\ohkimodmod{$\nu = 180$;} inset (a)] and \ohkimodmod{the} \ohkimodmod{(AA$'$-AA$'$)} symmetric-\ohkimodmod{edge pattern} [\ohkimodmod{$\nu = 179$;} inset (b)]. 
A clear change in the temperature dependence \ohkimodmod{appears at $T = T^*$ in $\left| d_{i_b=3 A \nu=180}(\pi / 2) \right|^2_{\rm Asym}$(open squares) and $\left| d_{i_b=30 A \nu=179}(\pi / 2) \right|^2_{\rm Sym}$ (filled circles)}.

For $T > T^*$ \ohkimodmod{emergent} edge states \ohkimodmod{are} clearly visible \ohkimodmod{for the} both \ohkimodmod{edge} patterns, \ohkimodmod{although} the \ohkimodmod{amplitude} of \ohkimodmod{$\left|d_{i_b=3 A \nu=180}(\pi / 2)\right|^2_{\rm Asym}$} is twice larger than \ohkimodmod{$\left|d_{i_b=3 A \nu=179}(\pi / 2)\right|^2_{\rm Sym}$}. 
Moreover, the edge state appears only at the right end \ohkimodmod{for} the \ohkimodmod{(AA$'$-BC) asymmetric case} \ohkimodmod{[inset (a)]}, whereas \ohkimodmod{it is} visible at both ends \ohkimodmod{for} the \ohkimodmod{(AA$'$-AA$'$) symmetric case} \ohkimodmod{[inset (b)]}. 
These results suggest that \ohkimodmod{the} electrons tend to gather more easily \ohkimodmod{around the} charge\ohkimodmod{-}neutral molecules \ohkimodmod{appearing} in the AA$'$ column \ohkimodmod{of} the edges (see Fig.~1).

For $T < T^*$ a redistribution of electrons takes place \ohkimodmod{for the (AA$'$-AA$'$) symmetric case,} \ohkimodmod{shifting} electron\ohkimodmod{s} from \ohkimodmod{an} edge (\ohkimodmod{$\left|d_{i_b A \nu=179}(\pi / 2)\right|^2_{\rm Sym}$} at $i_b = 3$) to the center (\ohkimodmod{$\left|d_{i_b A \nu=179}(\pi / 2)\right|^2_{\rm Sym}$} at $i_b = 30$) due to \ohkimodmod{a formation of a domain wall} and \ohkimodmod{an} associated bound state. 
\ohkimodmod{For} the \ohkimodmod{(AA$'$-BC)} asymmetric \ohkimodmod{case}, by contrast, electrons do not gather \ohkimodmod{around} the center, but instead the population around the right edge (\ohkimodmod{$\left|d_{i_b A \nu=180}(\pi / 2)\right|^2_{\rm Asym}$} at $i_b = 3$) suddenly increases below $T^*$, signaling that electrons are bound to \ohkimodmod{the} charge\ohkimodmod{-}neutral molecules \ohkimodmod{at that edge [see Fig.~1(a)]}. 
These edge-bound electrons might contribute to the conductivity by thermal excitations.

\begin{figure}
\begin{centering}
\includegraphics[width=80mm]{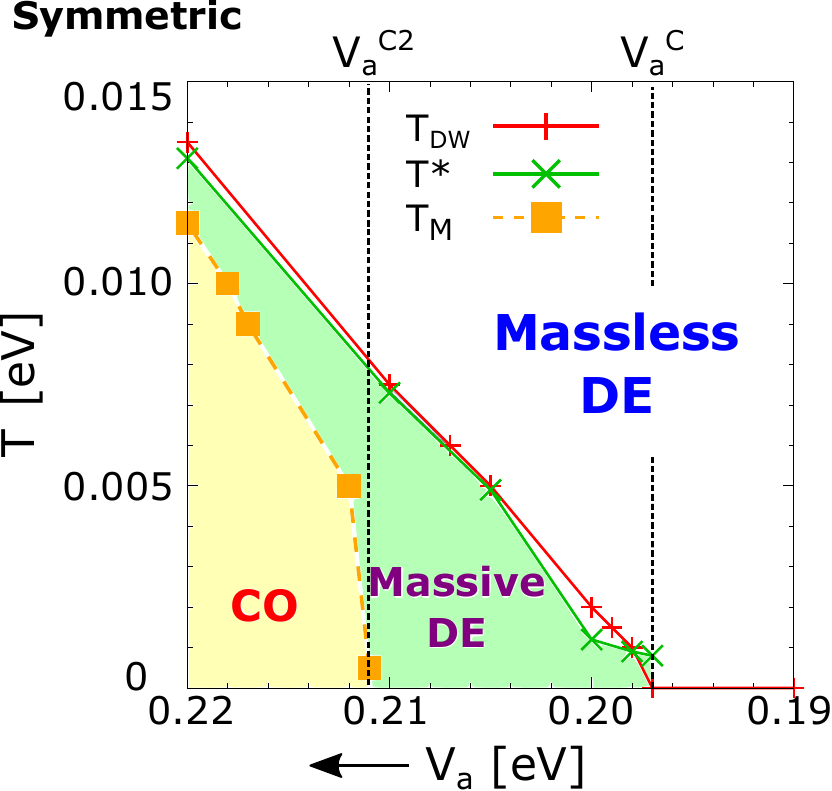}
\caption{\label{illustration}(Color online) 
The $V_a$-$T$ phase diagram \omori{for the} \ohkimodmod{(AA$'$-AA$'$)} \omori{symmetric-edge pattern, where \ohkimodmod{a} domain wall emerge}\ohkimodmod{s} \omori{in} \ohkimodmod{the} \omori{charge-ordered} \ohkimodmod{(CO)} \omori{phase}\ohkimodmod{s either with or without massive Dirac cones}. 
Gapless Dirac cones are present in the massless Dirac electron \ohkimodmod{(DE)} phase protected by the space and time inversion symmetry, whereas the inversion symmetry is broken in the \ohkimodmod{filled area} due to \ohkimodmod{the} charge-ordering. 
\omori{For} $T_{\rm M} < T < T^*$ (green region), a charge-ordering gap opens at the Dirac points and the cones become massive. 
These massive cones merge at $T = T_{\rm M}$ at the M point\ohkimodmod{. Bellow $T_{\rm M}$, a} charge-ordered state without \ohkimodmod{gapped} Dirac \ohkimodmod{cones is realized} (yellow region). 
}
\end{centering}
\end{figure}
 Figure 4 shows the \ohkimodmod{interaction-temperature (}$V_a$-$T$\ohkimodmod{)} phase diagram \ohkimodmod{for the \ohkimodmod{(AA$'$-AA$'$)} symmetric-edge pattern}, \ohkimodmod{where} \omori{we present three characteristic energy scales}\ohkimodmod{:} \omori{$T^*$, $T_{\rm DW}$, and $T_{\rm M}$.} 
\ohkimodmod{Already introduced} \omori{as} the \ohkimodmod{metal-to-insulator} transition temperature\ohkimodmod{, $T^*$} also \ohkimodmod{defines a phase boundary between the} massless \ohkimodmod{Dirac} phase (with gapless Dirac cones) and \ohkimodmod{the (charge-ordered)} massive \ohkimodmod{Dirac} phase (with gapped Dirac cones). 
\omori{$T_{\rm DW}$}\ohkimodmod{, which coincides with $T^*$, gives} \omori{the} energy scale for \ohkimodmod{forming a single domain wall} determined from \omori{the} \ohkimodmod{temperature where} the domain-wall width $W_{\rm D}$ \ohkimodmod{diverges} \omori{\cite{Matsuno2016, Ohki2018JPSJ}}\ohkimodmod{, and $T_{\rm M}$ represents} \omori{a} merging \ohkimodmod{transition} of \ohkimodmod{two} gapped Dirac cones\ohkimod{~\cite{Katayama2006, Kobayashi2007, Kobayashi2008IOP, Dietl, Montambaux2009PRB, Montambaux2009EPJB, Kobayashi2011, Matsuno2016, Omori2017}}. 
The fact that \ohkimodmod{we have $T^* = T_{\rm DW}$ directly} supports the notion that the domain wall \ohkimodmod{only} appears in the charge-ordered state and disappears in the massless Dirac phase. 
\ohkimodmod{To have a better understanding of this phase diagram, it is informative to focus on the evolution of the electronic state at low temperature as a function of $V_a$. 
As one increases $V_a$, there is first a transition from the massless \ohkimodmod{Dirac phase} to the massive Dirac phase at $V_a^c = 0.197$ (at $T = 0$) \ohkimodmod{occurring} simultaneously with the charge ordering~\cite{Matsuno2016, Omori2017, Ohki2018JPSJ}. 
Upon further increasing $V_a$, the merging transition takes place at $V_a^{c2} = 0.211$ (at $T=0.0005$) where the system changes from the (charge-ordered) massive Dirac state ($V_a^{c2} > V_a > V_a^c$) to the charge-ordered state with no Dirac cones ($V_a > V_a^{c2}$) (As we mentioned previously, the valley Chern number changes from a finite value to zero across this transition~\cite{Matsuno2016},} 
\ohkiadd{but in both phases metallic bound} \ohkimodmod{states exist along the} {domain wall}\ohkimodmod{). The latter critical interaction value well agrees with what is deduced from the previous study using the 2D periodic boundary condition ($V_a^{c2} = 0.212$ at $T = 0.001$~\cite{Ohki2018Crystals}).} 
\ohkimodmod{We note}, however, that the real situation is a bit more complicated \omori{because} the merging happens separately in the conduction and valence bands in $\alpha$-(BEDT-TTF)$_2$I$_3$ reflecting the tilt of the Dirac cones; \omori{in Fig.~4 we} \ohkimodmod{thus} defined $T_{\rm M}$ \ohkimodmod{from} the merging of two energy minima \ohkimodmod{taking place} in the conduction band\ohkimodmod{.}

\omori{
\subsection{Temperature Dependence of the Resistivity and the Transport Gap $\Delta_\rho$}
}

\begin{figure}
\begin{centering}
\includegraphics[width=75mm]{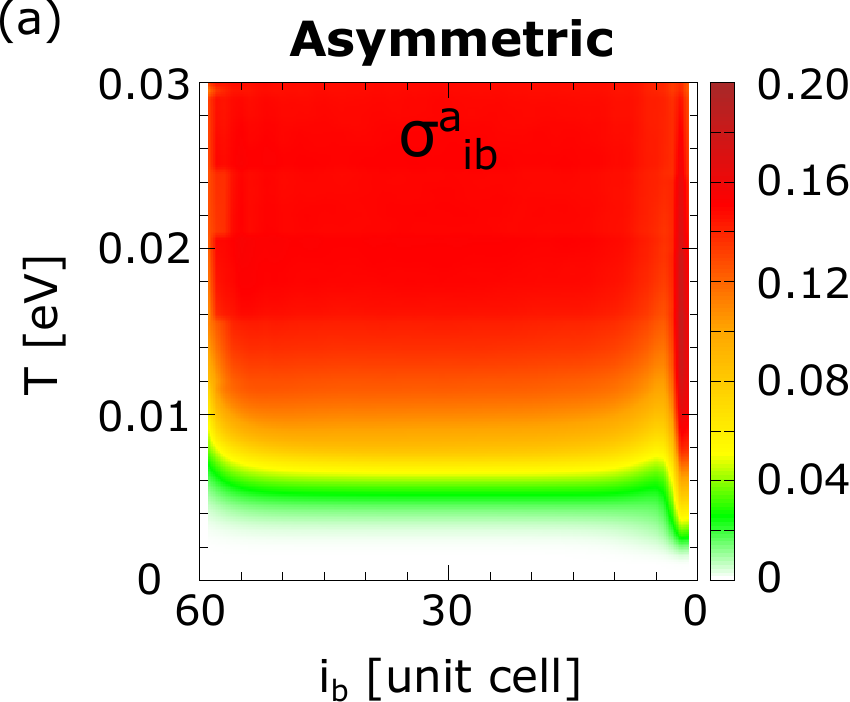}
\includegraphics[width=75mm]{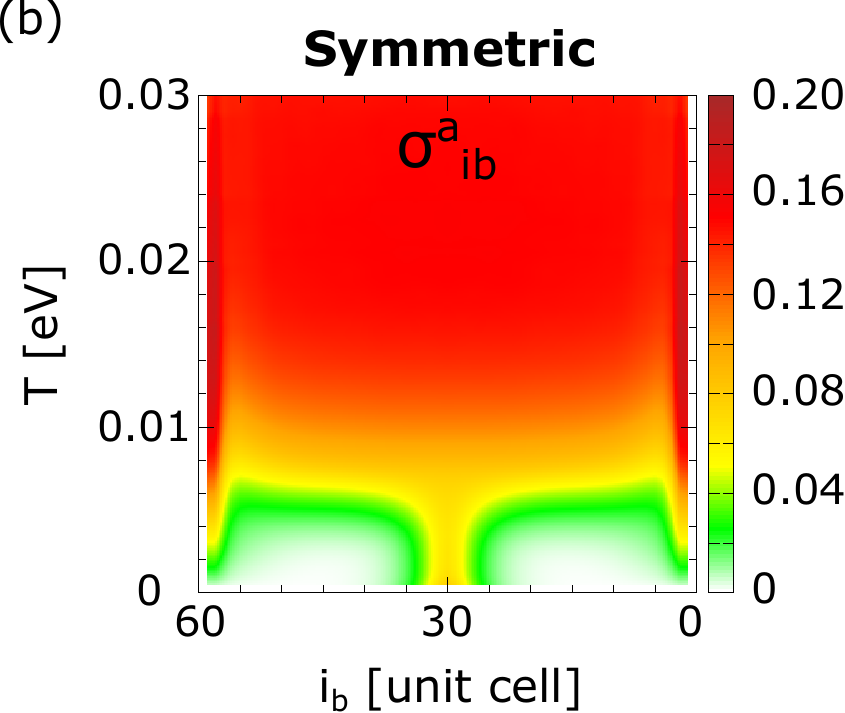}
\caption{\label{illustration}(Color online) 
2D plot of the spatially-resolved dc conductivity $\sigma^a_{i_b}$ in
 the \omori{$a$} direction at $V_a = 0.21$, plotted as a function of
 temperature $T$ and unit cell position in the \ohkimodmod{$b$} direction $i_b$ for (a) the (AA$'$-BC) asymmetric-edge pattern and (b) the (AA$'$-AA$'$) symmetric-edge pattern. 
}
\end{centering}
\end{figure}
%
\ohki{Next, we} \ohkimodmod{calculate} \omori{the temperature dependence of the resistivity to} \ohkimodmod{estimate} \omori{the transport gap $\Delta_\rho$.} 
Figures 5(a) and (b) present the 2D plots of the spatially resolved dc conductivity $\sigma^a_{i_b} (T)$ in the \omori{$a$} direction \ohkimodmod{(}at $V_a=0.21$\ohkimodmod{)} plotted as a function temperature \omori{$T$} and the position in the \omori{$b$} direction $i_b$ for the (AA$'$-BC) asymmetric-\omori{edge} \ohkimodmod{pattern} [Fig.~5(a)] and the (AA$'$-AA$'$) symmetric-\omori{edge} \ohkimodmod{pattern} [Fig.~5(b)]. 
The plots reveal a very different nature in the low $T$ behaviors \ohkimodmod{for two cases}. 
For the \ohkimodmod{(AA$'$-BC)} asymmetric pattern, $\sigma^a_{i_b} (T)$ vanishes at low $T$ in the bulk of the sample since a gap opens at the Fermi energy. 
For the \ohkimodmod{(AA$'$-AA$'$)} symmetric pattern, by contrast, $\sigma^a_{i_b} (T)$ becomes vanishingly small at low $T$ except for the \ohkimodmod{central} region of the sample (i.e., $i_b = 30$) where a gapless 1D bound state appears on the domain wall \omori{[Fig.~2(b)]} \ohkimodmod{and yields} finite conductivity~\omori{\cite{Omori2017}}. 
Note that high conductivity also survives at low $T$ on the two edges in the \ohkimodmod{(AA$'$-AA$'$)} symmetric case, \ohkimodmod{whereas} it remains large \ohkimodmod{only} on the right end in the \ohkimodmod{(AA$'$-BC)} asymmetric case \ohkimodmod{(\omori{for $T \gtrsim 0.002$}).} 
\ohkimodmod{These high conductivities} can be explained by thermal edge conductance \ohkimodmod{owing to the} edge-\ohkimodmod{bound electrons, as we mentioned in the previous section} (see Figs.~2 and 3).

\begin{figure}
\begin{centering}
\includegraphics[width=86mm]{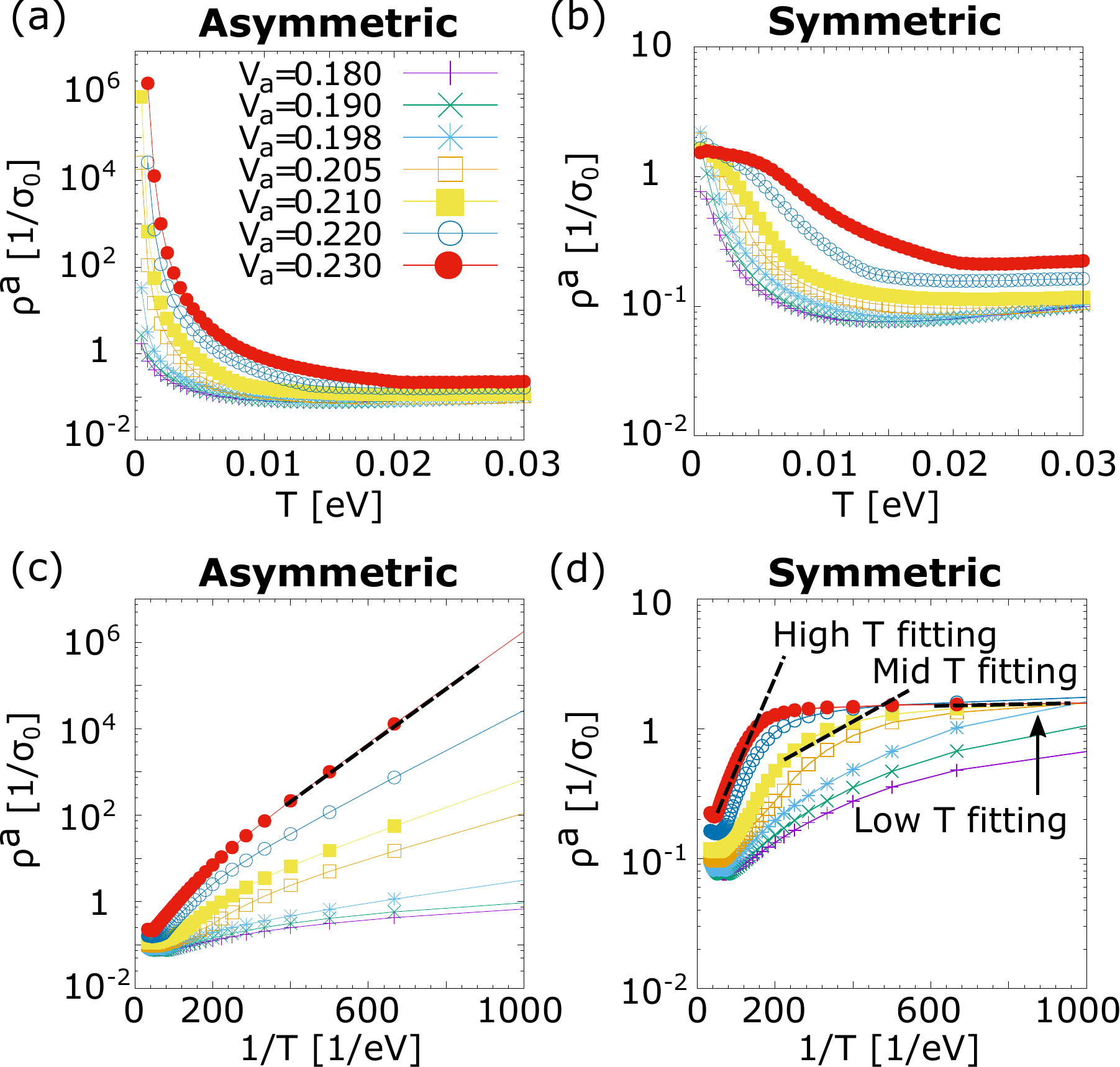}
\caption{\label{illustration}(Color online) 
Temperature dependence of the dc resistivity $\rho^a$ in the \ohkimodmod{$a$} direction
 plotted for several values of $V_a$ for (a) \ohkimodmod{(AA$'$-BC)} asymmetric-edge and (b) \ohkimodmod{(AA$'$-AA$'$)} symmetric-edge patterns. 
The corresponding Arrhenius plots of (a) and (b) are presented in (c) and (d), respectively. 
Representative fits to the data are shown by black dashed lines. 
}
\end{centering}
\end{figure}
To have a closer look on the temperature dependence, we measured the dc resistivity $\rho^a (T)$ as a function of $T$ in the $a$ direction for several strength\ohkimodmod{s} of \omori{interaction $V_a$} ranging from 0.180 to \omori{0.230}. 
The resulting curves for the \ohkimodmod{(AA$'$-BC)} \omori{asymmetric}- and \ohkimodmod{(AA$'$-AA$'$)} \omori{symmetric-edge} patterns are shown in \omori{Figs.}~6(a) and 6(b), respectively. 
For the \ohkimodmod{(AA$'$-BC)} asymmetric pattern [Fig.~6(a)], the curves start to diverge at low $T$ above \ohkimodmod{the} critical interaction $V_a^c = 0.197$. 
\omori{This divergence} signals the \ohkimodmod{opening} of a gap via spontaneous charge ordering, in line with our recent results using the \ohkimodmod{2D} periodic boundary condition~\cite{Ohki2018Crystals}. 
These data are contrasted to the \ohkimodmod{(AA$'$-AA$'$)} symmetric pattern [Fig.~6(b)] in which the resistivity does not diverge but levels off at low $T$ for $V_a > V_a^c$ since the \omori{domain-wall} formation in the charge-ordered state results in \ohkimodmod{a} finite conductivity along the gapless bound state \omori{[Fig.~5(b)]}. 
The values of the resistivity gap $2\Delta_{\rm \rho}$ can be extracted from exponential fits to the Arrhenius plot of $\rho^a(1/T)$, presented in Figs.~6(c) and 6(d) for the \ohkimodmod{(AA$'$-BC)} asymmetric- and \ohkimodmod{(AA$'$-AA$'$)} symmetric\ohkimodmod{-edge} patterns, respectively. 
\ohkimodmod{Upon increasing $V_a$, the} slope of the data shows a systematic increase for $V_a \ge V_a^c$ \ohkimodmod{in} the \ohkimodmod{(AA$'$-BC)} asymmetric \ohkimodmod{case} [Fig.~6(c)], pointing to a continuous evolution of $2\Delta_{\rm \rho}$ as \ohkimodmod{a function of} $V_a$. 
For the \ohkimodmod{(AA$'$-AA$'$)} symmetric \ohkimodmod{case} [Fig.~6(d)], however, the curves of $\rho^a(1/T)$ show a nonmonotonic behavior \ohkimodmod{as one increases} $V_a$; for large values of $V_a$ there is a steep slope at high $T$ (around the transition temperature $T^*$), a gradual increase at intermediate $T$ ($\simeq T^*/2$), and a clear levelling-off at low $T$ ($\simeq 0.001 \omori{\ll} T^*$). 
These contrasting behaviors yield different estimates of $2\Delta_{\rm \rho}$ in the charge-ordered state ($V_a > V_a^c$) that strongly depend on the range of the fits to the data, as we shall see below.

\begin{figure}
\begin{centering}
\includegraphics[width=80mm]{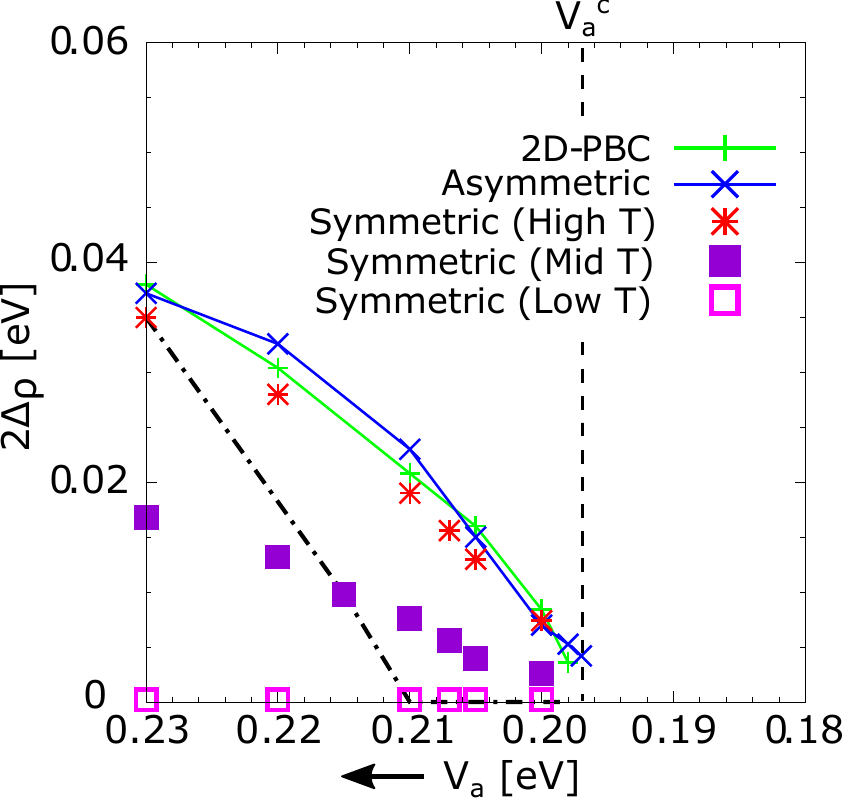}
\caption{\label{illustration}(Color online) 
The resistivity gap $2\Delta_\rho$ plotted as a function of $V_a$ for
 the 2D periodic boundary condition (pluses)~\omori{\cite{Ohki2018Crystals}}, the \ohkimodmod{(AA$'$-BC)} \omori{asymmetric-edge} \ohkimodmod{pattern} (crosses), \ohkimodmod{and} the \ohkimodmod{(AA$'$-AA$'$)} \omori{symmetric-edge} \ohkimodmod{pattern. For the last one, results obtained for three distinct fitting regimes are presented:} high $T$ \ohkimodmod{fits} (stars), mid $T$ \ohkimodmod{fits} (filled squares), and low $T$ \ohkimodmod{fits} (open squares). 
$V_a^c$ stands for the critical point in this model at which the charge-ordering gap closes. 
The black dash-dotted line corresponds to the $V_a$ dependence of $2\Delta_\rho$ that is deduced from the experimental data in Ref.~\cite{Liu2016} following the fitting procedure discussed in the text. 
}
\end{centering}
\end{figure}

%
We note here that the calculated \ohkimodmod{curves of} $\rho^a(1/T)$ in Fig.~6(d) have many features in common with the recent \ohkimodmod{experimental data} at a range of pressure \omori{$P$}~\cite{Liu2016}, 
\ohkimodmod{showing} strongly nonmonotonic changes in the shape of $\rho^a(1/T)$ upon increasing $P$. 
This apparent similarity motivates us to make a comparison of the calculated values of $2\Delta_{\rm \rho} (V_a)$ with the experimentally obtained $2\Delta_{\rm \rho} (P)$, \omori{considering $V_a$ instead of $P$ as} \ohkimodmod{a} \omori{control parameter of the charge-order}\ohkimodmod{ing} \omori{transition.} 
To this end, we \omori{perform} \ohkimodmod{fits using} \omori{different} \ohkimodmod{fitting ranges to} Figs.~6(c) and 6(d). 
Figure 7 plots the resulting values of $2\Delta_{\rm \rho}$ against $V_a$ for several fitting procedures. 
\omori{The} horizontal axis is reverted to make \ohkimodmod{a} comparison to the pressure-dependence data in Ref.~\cite{Liu2016}. 
The charge-ordering gap $2\Delta_{\rm \rho} (V_a)$ starts to open at $V_a = V_a^c$ and continues to develop almost linearly for $V_a > V_a^c$. 
\omori{\ohkimodmod{For} the \ohkimodmod{(AA$'$-BC)} asymmetric pattern \ohkimodmod{the gap size} agrees with} \ohkimodmod{that} for the 2D \ohkimodmod{periodic} boundary condition~\omori{\cite{Ohki2018Crystals}} since the bulk appearance of charge order results in a unique definition of the gap. 
In contrast, the evolution of $2\Delta_{\rm \rho} (V_a)$ \omori{for the \ohkimodmod{(AA$'$-AA$'$)} symmetric} \ohkiadd{pattern} shows a marked difference \ohkimodmod{for certain types of fits}. 
\omori{The} fits at high $T$ \ohkiadd{yield} \ohkimodmod{a similar result to} the above two cases, pointing to a small influence of the \ohkimodmod{domain-wall} bound state at high temperature. 
On the other hand, the gap becomes smaller for the intermediate-$T$ \ohkimodmod{fits} and is almost zero for the low-$T$ \ohkimodmod{fits}, reflecting the presence of the metallic bound state \ohkimodmod{dominating} the conductivity at low temperature.

In Ref.~\cite{Liu2016} the authors examined fits to the Arrhenius plot of experimental resistivity at various values of $P$ for determining $2\Delta_{\rm \rho} (P)$, but their choice of fitted temperature range \ohkimodmod{varies from pressure to pressure: 
(i) At} \omori{ambient pressure} \ohkimodmod{the fit was performed} \omori{just below the transition temperature (High $T$), (ii) \ohkimodmod{for} $P = 4.8$ and $6.3$ kbar} \ohkimodmod{it was} \omori{done at half of the transition temperature (Mid $T$), and (iii) for $7$ kbar $< P < P_c$ \ohkimodmod{it was done} at \ohkimodmod{a} low-temperature limit below 20 K (Low $T$).} 
\ohkimodmod{Given} the similarity of the experimental Arrhenius plot and Fig.~6 (d), one is
tempted to hypothesize that there is a correspondence between the
$P$-dependent choice of the \omori{fitted temperature range} \ohkimodmod{in~\cite{Liu2016}} and the $V_a$-dependent fit range in Fig.~6(d). 
By doing so, we are able to estimate an expected curve of $2\Delta_{\rm \rho} (V_a)$ \ohkimodmod{corresponding to the experimental data} \ohkimodmod{(\omori{indicated by the black dash-dotted line in Fig.~7})}. 
\\
\\
\\

\subsection{Optical Conductivity \omori{and Optical gap $\Delta_{\rm O}$}}
%

\begin{figure}
\begin{centering}
\includegraphics[width=85mm]{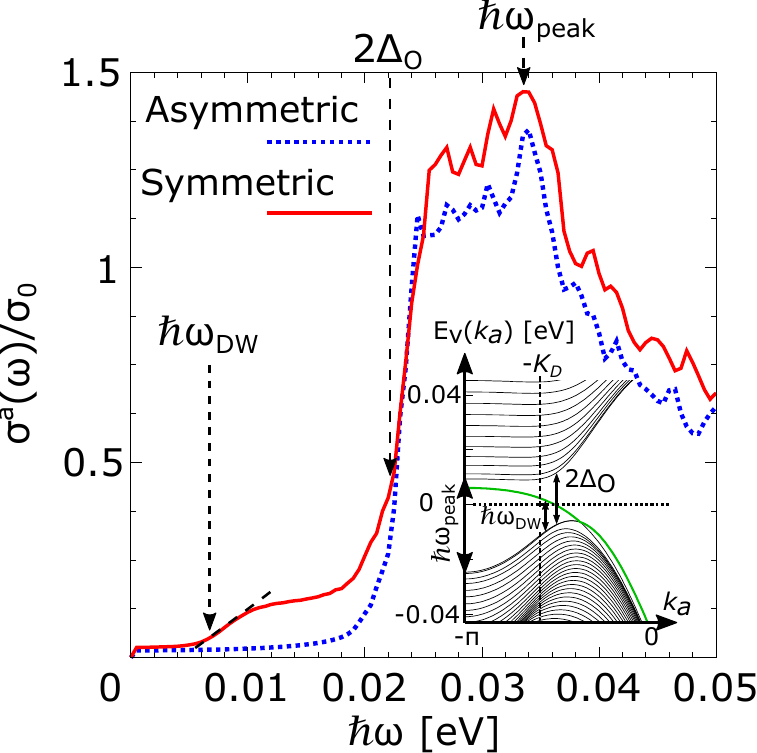}
\caption{\label{illustration}(Color online) 
Optical conductivity spectra relative to the universal conductivity $\sigma^a(\omega)/\sigma_0$, plotted as a function of energy $\omega$ at $V_a = 0.21$ and $T = 0.0005$ for the \ohkimodmod{(AA$'$-BC)} \omori{asymmetric-edge} (dashed curve) and the \ohkimodmod{(AA$'$-AA$'$)} \omori{symmetric-edge} (solid curve). 
\ohki{The contribution \ohkimodmod{from} the Drude term is subtracted.} 
$\hbar \omega_{\rm peak}$ is given by the maximum energy of the spectra, \omori{and} the optical gap $2\Delta_{\rm O}$ is defined as the inflection point. 
The black dashed line at low energy is drawn to determine $\hbar \omega_{\rm DW}$. 
Inset: The energy bands near the Fermi energy for the \ohkimodmod{(AA$'$-AA$'$)} symmetric-edge pattern. 
The gapless 1D bound state stemming from the \omori{domain wall} is
 shown by the green solid curve intersecting the Fermi energy \ohkiadd{near the \ohkimodmod{gapped} Dirac point} at \omori{$k_a = -K_{\rm D}$. }
The corresponding energy scales of $\hbar \omega_{\rm peak}$, $\hbar \omega_{\rm DW}$ and $2\Delta_{\rm O}$ are indicated by arrows. 
}
\end{centering}
\end{figure}
As \ohkimodmod{a} next step, we calculated the optical conductivity $\sigma^a(\omega)$ in the \omori{$a$} direction for the \ohkimodmod{(AA$'$-BC)} asymmetric- and \ohkimodmod{(AA$'$-AA$'$)} symmetric-edge \ohkimodmod{patterns, where the} interaction $V_a$ was varied between 0.180 and \omori{0.230}. 
The representative conductivity data normalized to $\sigma_0 = 4e^2/\pi
h$ deep in the charge-ordered phase \ohkimodmod{(}\omori{at $T = 0.0005$ and $V_a=0.21$}\ohkimodmod{)} are presented in Fig.~8.
The overall shape of the optical conductivity spectra is more or less
similar for the two edge patterns \ohkimodmod{and} is characterized by \omori{two energy scales, $2\Delta_{\rm O}$ and $\hbar \omega_{\rm peak}$. A} sharp drop at $2\Delta_{\rm O} \simeq 20$ meV \omori{signals} the
opening of a direct charge-ordering gap at the Dirac point. \omori{Above
the gap, there is}
a hump-like structure with a peak at $\hbar \omega_{\rm peak} \simeq 34$ meV.
The peak is ascribed to a direct transition between different van Hove singularities in the conduction and valence bands, locating at a time reversal invariant momentum (TRIM). 
Similar structures have been observed in \ohkimodmod{the} previous study using the 2D periodic boundary condition~\cite{Ohki2018Crystals}. 
\ohkimodmod{A} remarkable \omori{difference \ohkimodmod{found} in the present cylindrical model is} \ohkimodmod{the} additional bump at low energy with a kink at $\hbar \omega_{\rm DW} \simeq 6$ meV, which only appears \ohkimodmod{for} the \ohkimodmod{(AA$'$-AA$'$)} symmetric pattern. 
This bump can be associated to a direct transition between the \ohkimod{valence} band and the gapless band \ohkimodmod{at the Fermi energy linked to} the 1D bound state on the domain wall. In addition, we note that $\hbar \omega_{\rm DW}$ is almost half the charge-ordering gap $2\Delta_{\rm O}$ since the Fermi energy locates at approximately the mid point of the gap (for details, see the inset of Fig.~8).

\begin{figure}
\begin{centering}
\includegraphics[width=75mm]{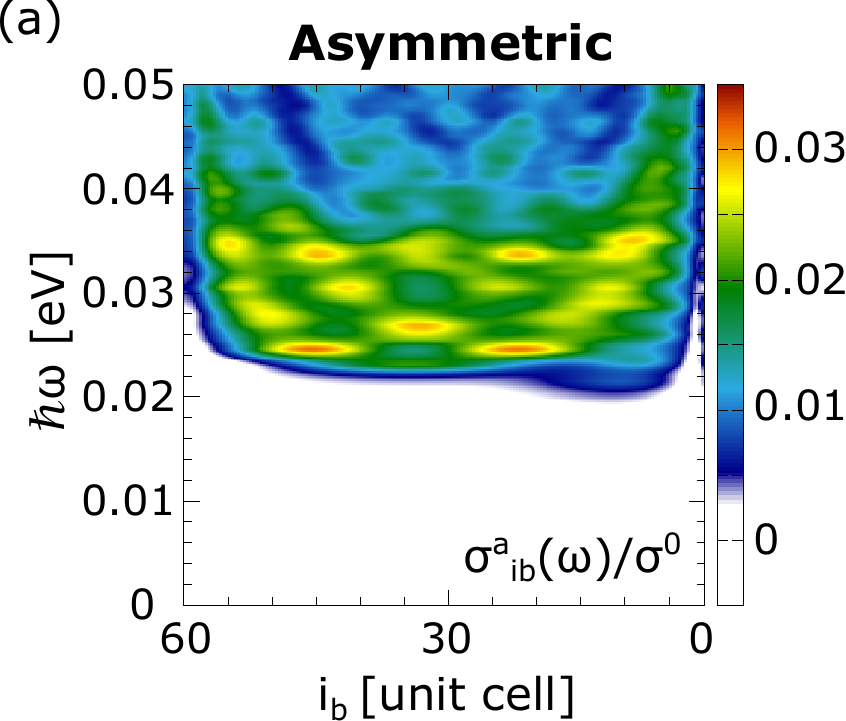}
\includegraphics[width=75mm]{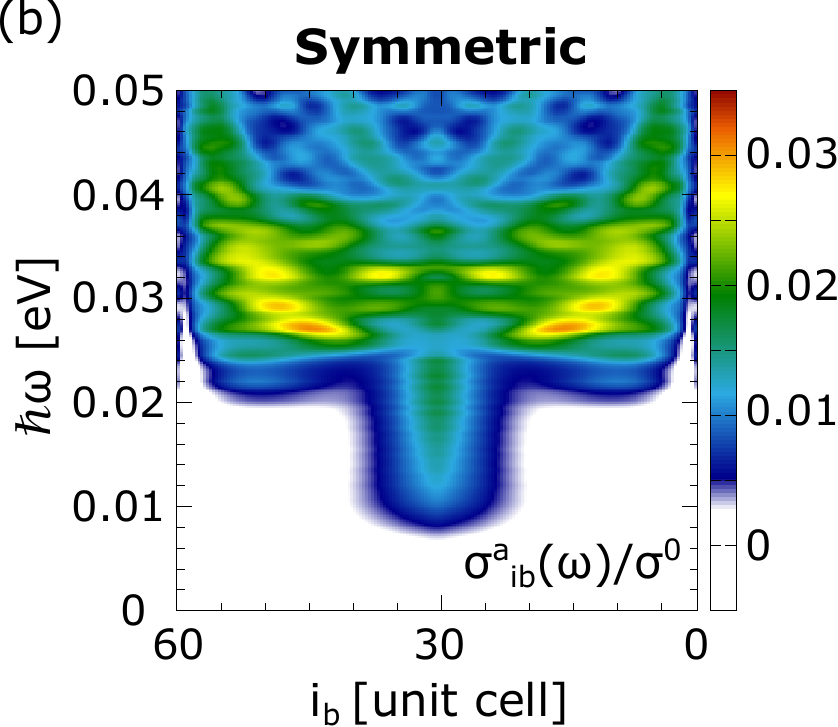}
\caption{\label{illustration}(Color online) 
The 2D plot of the spatially-resolved optical conductivity
 $\sigma^a_{i_b}(\omega)/\sigma_0$ at $V_a = 0.21$ and $T = 0.0005$, plotted as a function of energy $\hbar \omega$ and the unit cell position $i_b$ \ohkimodmod{in the $b$ direction} for the \ohkimodmod{(AA$'$-BC)} asymmetric-edge pattern (a) and the \ohkimodmod{(AA$'$-AA$'$)} symmetric-edge pattern (b). 
}
\end{centering}
\end{figure}
Figures 9(a) and 9(b) show the 2D plots of the spatially resolved optical conductivity $\sigma^a_{i_b}(\omega)$ normalized to $\sigma_0$
in the charge-ordered phase \ohkimodmod{(}\omori{at $T = 0.0005$}\ohkimodmod{)} calculated at $V_a = 0.21$ for the \ohkimodmod{(AA$'$-BC)} \omori{asymmetric-edge} \ohkimodmod{pattern} [Fig.~9(a)] and the \ohkimodmod{(AA$'$-AA$'$)} \omori{symmetric-edge} \ohkimodmod{pattern} [Fig.~9(b)], plotted as a function of $\hbar \omega$ and the position $i_b$ in the \omori{$b$} direction. 
A spatially uniform optical gap is clearly visible in both plots except for the central region ($i_b \sim 30$) in \ohkimodmod{Fig.~9(b) where} the conductivity increases from a lower energy due to the \ohkimodmod{domain-wall} bound state, bringing about a T-shaped structure. 
At the edges finite conductivity resumes due to \ohkimodmod{a} direct transition between the edge states (Fig.~3) and the conduction band. 
The edge conductance is absent at the left \omori{edge} ($i_b = 60$) \ohkimodmod{for} \omori{the \ohkimodmod{(AA$'$-BC)} asymmetric pattern} in Fig.~9(a) since there is no edge states in this \ohkimodmod{case}.

\begin{figure}
\begin{centering}
\includegraphics[width=80mm]{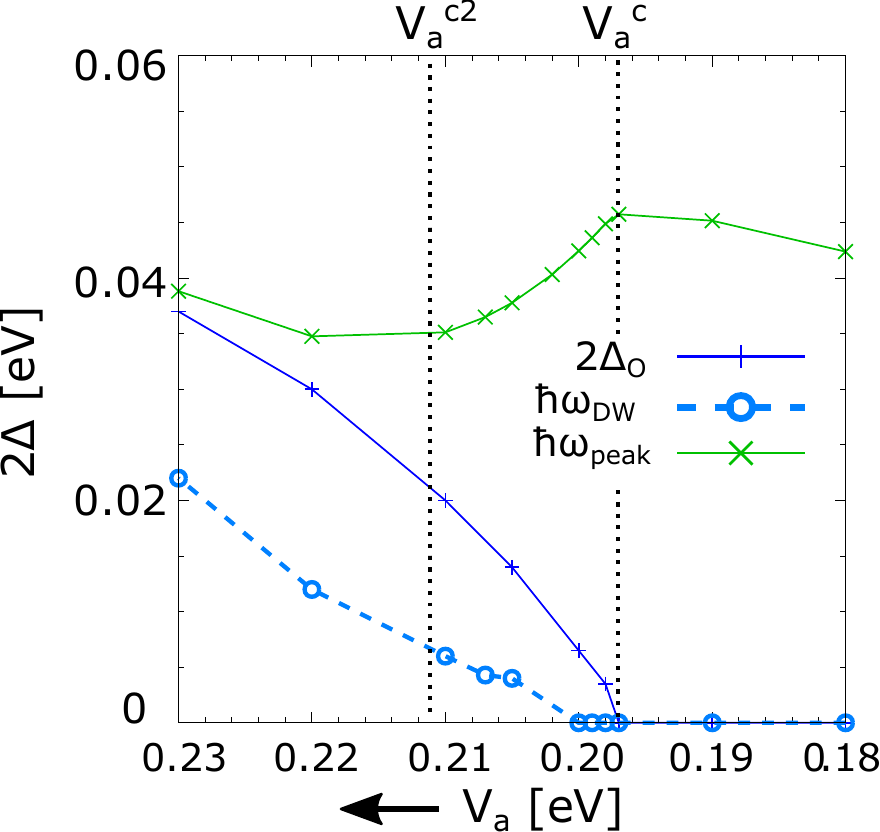}
\caption{\label{illustration}(Color online) 
The optical gap $2\Delta_{\rm O}$ plotted against $V_a$ (crosses). 
Other characteristic energy scales of $\hbar \omega_{\rm peak}$ (stars) and $\hbar \omega_{\rm DW}$ (open circles) are also presented. 
Here, $\hbar \omega_{\rm peak}$ \ohkimodmod{corresponds to} the energy difference between the conduction and valence bands at the M-point in the first Brillouin zone. 
\ohkimodmod{Note that the results are the same for both edge types.} 
}
\end{centering}
\end{figure}
In Fig.~10 we present results for the optical gap $2\Delta_{\rm O}$ and the two characteristic energy scales, $\hbar \omega_{\rm peak}$ and $\hbar \omega_{\rm DW}$ 
\ohkimodmod{(The} horizontal axis is inverted \ohkimodmod{to make a} comparison with the transport gaps in Fig.~7\ohkimodmod{)}. 
The results of $2\Delta_{\rm O}$ and $\hbar \omega_{\rm peak}$ show identical behaviors for the \ohkimodmod{(AA$'$-BC)} asymmetric and \ohkimodmod{(AA$'$-AA$'$)} symmetric patterns and will not be distinguished. 
The optical gap $2\Delta_{\rm O}$ approximately increases linearly with increasing $V_a$, which is paralleled by $\hbar \omega_{\rm DW}$ that also linearly develops in a similar fashion\ohkimodmod{. Note that} \omori{the magnitude of} \ohkimodmod{$\hbar \omega_{\rm DW}$ is about 50 $\%$ the size of} \omori{$2\Delta_{\rm O}$} as we mentioned above. 
\ohkimod{By contrast,} \ohkimodmod{$\hbar \omega_{\rm peak}$ is \ohkiadd{almost} flat at lower $V_a$, }\ohkimodmod{starts to} \ohkiadd{decrease \ohkimodmod{at} $V_a = V_a^c$ (due to a deformation of \ohkimodmod{energy bands at the} charge-ordering transition~\cite{Ohki2018Crystals})}\ohkimodmod{, gradually approaches} $2\Delta_{\rm O}$ for $V_a^{c2} > V_a > V_a^c$, and \ohkimodmod{eventually} becomes almost identical to $2\Delta_{\rm O}$ for $V_a > V_a^{c2}$. 
The last point agrees with the fact that the van Hove singularities in the conduction and valence bands disappear when the merging transition \ohkimodmod{occurs} \omori{at $V_a = V_a^{c2}$} \ohkimodmod{since} \ohkimodmod{in that case} direct transitions between the singular points also vanish. 
We have reported similar asymptotic behaviors of $2\Delta_{\rm O}$ and $\hbar \omega_{\rm peak}$ for the 2D periodic boundary condition in Ref.~\cite{Ohki2018Crystals}. 
\omori{This similarity suggests} that the overall changes of the band topology \omori{and the optical gap are} irrespective of boundary choices (\omori{i.e., the \ohkimodmod{absence or presence} of \ohkimodmod{a} domain wall}).

\section{Summary and discussion}

%
\ohkimod{In this study} \ohkimodmod{we examined the detailed temperature dependence of the electronic states of interacting 2D Dirac electrons in $\alpha$-(BEDT-TTF)$_2$I$_3$ by using a minimal lattice model at a mean-field level. 
More specifically, we assumed cylindrical boundary conditions that, depending on the type of edges, necessitate a single domain wall as a consequence of geometrical constraints.} 
\ohkimodmod{An} \ohkimod{interaction-temperature ($V_a$-$T$) phase diagram} \ohkimodmod{is proposed} \ohkimod{which} \ohkimodmod{we claim to correspond} \ohkimod{to the} \ohkimodmod{experimentally-reported} \ohkimod{pressure-temperature ($P$-$T$) phase diagram} \ohkimodmod{if one associates an increase in $P$ with a reduction in $V_a$.}

\ohkimodmod{
We find clear evidence for a formation of a domain wall in the entire charge-ordered phase ($V_a > V_a^c$) for the (AA$'$-AA$'$) symmetric-edge pattern. 
The key finding is the discovery of a conduction mechanism along the 1D domain-wall bound state that excellently explains the experimental discrepancy in the sizes of transport and optical gaps under pressure. 
The conduction along this bound state turns out to offer a major contribution to the transport conductivity at large $V_a$, whose degree increases upon cooling and completely dominates the conduction at low temperature, resulting in a saturation of resistivity and a vanishingly small transport gap. 
In contrast, the impact of the bound state is rather limited in the optical conductivity, giving rise to a finite gap in the optical spectra regardless of the size of $V_a$ in the charge-ordered phase [The bound state causes, however, a small additional bump at low energy ($\hbar \omega_{\rm DW}$) corresponding to an in-gap bound state]. 
As a direct consequence of these differences, the transport gap ends up with a vanishingly small value at low temperature, whereas the optical gap stays finite, in a remarkable agreement with the experiments. 
We therefore conclude that the emergent domain wall must be relevant to the observed discrepancy in the experimental gap sizes. 
This point is in line with our supportive calculations using the (AA$'$-BC) asymmetric-edge pattern and the 2D periodic boundary condition since in these cases there is no domain wall, and the two gaps reasonably coincide. 
}

\ohkimodmod{We also demonstrate that} the Arrhenius analyses of the resistivity data provide unusual gap values \ohkimodmod{for large $V_a$} depending strongly on the $T$ slice the data are fitted. 
The gap is identical to the bulk gap just below the transition temperature $T^*$, whereas it is largely suppressed towards lower $T$ and eventually becomes zero at low $T$. 
We reiterate that a very similar behavior has been reported in the recent transport experiments under pressure; Fig.~1(b) of Ref.~\cite{Liu2016} highlights at low $T$ a levelling-off-like feature of resistivity developing upon increasing pressure, which draws an excellent parallel with the calculated resistivity in Fig.~6(d) showing a similar saturation at large $V_a$. 
This strongly reinforces our original hypothesis that the major impact of pressurization is altering the size of $V_a$.

Of course \ohkimodmod{our} assumption in \ohkimodmod{this} model is \ohkimodmod{somewhat} oversimplified, and in reality pressure would also changes other parameters such as the electronic bandwidth as well as interactions between BEDT-TTF molecules and I$_3$ anions~\cite{Alemany}. 
In our view, however, these effects are insufficient to qualitatively reproduce the remarkable saturation of low-$T$ resistivity, allowing us to safely omit \ohkimodmod{these effect} as a first approximation. 
Unfortunately, the putative model we rely on precludes us from making a more quantitative analysis of the gap sizes at this stage. 
In this regard a more complete calculation considering all these pressurization effects may prove interesting.

\ohkimodmod{
The finding of the phase diagram also reveals that the charge-ordered phase is divided into two subgroups in an extended region on the $V_a$-$T$ plane; namely, the massive Dirac electron phase with gapped Dirac cones on the lower-interaction side ($V_a^{c2} > V_a > V_a^c$) and the trivial charge-ordered phase without any cone on the higher-interaction side ($V_a > V_a^{c2}$). The categorization of these two phases in terms of the valley Hall effect would be informative for improving further understanding of charge order in this material.
}

\ohkimodmod{We note} that while our model assumes semi-infinite boundary conditions and is hence different from the situation in naturally grown $\alpha$-(BEDT-TTF)$_2$I$_3$ bulk samples, a real crystal is a quasi-2D material consisted of a multiple stacks of 2D conducting layers which inevitably has edges in each layer of either \ohkimodmod{(AA$'$-BC)} asymmetric- or \ohkimodmod{(AA$'$-AA$'$)} symmetric-edge patterns. 
Taking proper account of this\ohkimodmod{, one can} expect \ohkimodmod{that at least} certain portions of the layers \ohkimodmod{are} comprised of the \ohkimodmod{(AA$'$-AA$'$)} symmetric-edge pattern. 
We therefore argue that \ohkimodmod{the} domain-wall conductance must be relevant at low temperature as long as these \ohkimodmod{symmetrically} edged layers are concerned. 
Recent experimental reports on the electronic ferroelectricity in $\alpha$-(BEDT-TTF)$_2$I$_3$ at ambient pressure~\cite{Yamamoto2008,Yamamoto2010,Lunkenheimer} agree with this notion, which point to the presence of multiple domain walls created between charge-ordered regions having opposing electric polarizations. 
Another interesting remark is that the saturation-like behavior of resistivity emerges even at ambient pressure in some samples~\cite{Ishikawa2016} (albeit it is absent in others~\cite{Liu2016, Ivek2017}). 
For real materials this suggests that the domain wall conduction may be present in a much wider region in the charge-ordered phase. 
We propose that optical conductivity or real-space resolved spectroscopy would be able to confirm domain walls, which should see some \ohkimodmod{bump} structures at low energy inside the charge-ordering gap.

\ohkimodmod{
\ohkimodmod{Finally}, let us recall that our mean-field theory \ohkimodmod{assumes} a minimal model which only \ohkimodmod{considers} a single domain wall necessitated by geometrical requirements (i.e., the type of edges at the two ends in the $b$ direction). 
To go one step ahead, one should take, for instance, thermal effects into account which can cause excitations of multiple domain walls by fluctuations. 
Nevertheless, we believe that the qualitative features of the transport and optical gaps, in particular their discrepancy in size, would not be altered much by the presence of excess domain walls. 
In that regard, our arguments are expected to be durable even in the presence of fluctuations, although further studies are clearly needed for gaining deeper insight into these phenomena. 
It will be an interesting future problem to consider the creation and annihilation of domain walls as well as their dynamic properties. 
}
\\
\\
\\
\\
\\

\begin{acknowledgments}
The authors would like to thank S. Onari, K. Miyagawa, K. Kanoda, and H. Fukuyama for fruitful discussions, \ohkimodmod{and} Y. Yamakawa for advice on numerical calculation\ohkiadd{. We \ohkimodmod{would also} like to express our \ohkimodmod{gratitude} to} \ohkimodmod{M. Hirata for discussions as well as reading and correcting the manuscript.}
This work was supported by MEXT/JSPJ KAKENHI under Grant Nos. 15K05166 \ohkiadd{and 19J20677}.
\end{acknowledgments}

\appendix
\renewcommand{\thefigure}{\Alph{section}.\arabic{figure}}
\setcounter{figure}{0}
\section{\ohki{Details of the formulation in the cylindrical systems using extended Hubbard model}}

%
 \ohki{In Appendix A we introduce the \ohkimodmod{detailed} formulations used in \ohkimodmod{this study}. Our starting point is \ohkimodmod{a} Hamiltonian which is \ohkimodmod{described in the} site representation \ohkimodmod{[see equations (1) and (2) in the main text]}.}  
\ohkimodmod{Making a} Fourier transformation along the $a$ axis, \ohkimodmod{we define the Fourier \ohkiadd{inverse} transform} ${a_{l \sigma}}={N_a^{-1/2}{\sum_{k_a}}a_{k_a l_b \sigma}e^{ik_a i_a}}$ for \ohkimodmod{site} $l_b = (i_b, \alpha)$ \ohkimodmod{($i_b$-th unit cell in the $b$ direction and $\alpha$-th molecule)} and wavenumber vector $k_a$\ohkimodmod{. The} Hamiltonian \ohkimodmod{in the wavenumber representation then reads} 
\begin{eqnarray}
{H_{\rm kin}} &=&  {\sum_{l_b {l_b}'}} {\sum_{k_a \sigma}} {\varepsilon}_{l_b, {l_b}'}(k_a)
a^{\dag}_{k_a l_b \sigma}a_{k_a {l_b}' \sigma} ,
\\
{H_{\rm int}} &=& \frac{U}{N_a} \sum_{l_b, {l_b}'} \sum_{k_a k_a' q_a} \delta_{l_b,{l_b}'}a^{\dag}_{k_a - q_a l_b \uparrow} a^{\dag}_{k_a' + q_a {l_b}' \downarrow} a_{k_a' {l_b}' \downarrow}a_{k_a l_b \uparrow} \nonumber\\
&&+\frac{1}{N_a}\sum_{l_b {l_b}'} \sum_{k_a k_a' q_a}\sum_{\sigma \sigma'}V_{l_b, {l_b}'}(q_a) \nonumber\\
&&\hspace{1cm} \times a^{\dag}_{k_a-q_a l_b {\sigma}}a^{\dag}_{k_a'+q_a {l_b}' \sigma'}a_{k_a' {l_b}' \sigma'}a_{k_a l_b \sigma} \nonumber\\
&&+\sum_{edge}\sum_{k_a}V_{\rm edge}a^{\dag}_{k_a l_b}a_{k_a l_b}.
\end{eqnarray}
\ohkimodmod{Here,} ${\varepsilon}_{l_b, {l_b}'}(k_a)$ and $V_{l_b, l_b'}(q_a)$ are defined as ${\varepsilon}_{l_b, {l_b}'}(k_a) = {\sum_{\delta}} t_{l_b, {l_b}'}e^{-ik_a{\delta}}$ and $V_{l_b, {l_b}'}(q_a) = \frac{1}{2} \sum_{\delta}V_{l_b, l_b'}$, respectively, \ohkimodmod{with} a momentum transfer $q_a$ and a vector $\delta$ connecting all possible-nearest neighbor sites. 
Within the Hartree approximation one obtains a mean-field Hamiltonian 
\begin{eqnarray}
\hspace{-0.7cm}{H_{\rm MF}} &=& {\sum_{l_b, {l_b}'}} {\sum_{k_a \sigma}} \tilde{\varepsilon}_{l_b {l_b}' {\sigma}}(k_a)a^{\dag}_{k_a l_b \sigma}a_{k_a {l_b}' \sigma}{\rm +const.} ,\\
\hspace{-0.7cm}{\tilde{\varepsilon}_{l_b {l_b}' \sigma}(k_a)} &=& \varepsilon_{l_b, {l_b}'}(k_a) + {\delta}_{l_b l_b'} \phi_{l_b \sigma} ,\\
\hspace{-0.7cm}\phi_{l_b \sigma} &=& U\langle{n_{l_b -\sigma}}{\rangle} + {\sum_{{l_b}'' \sigma'}}V_{l_b, {l_b}''}{\langle}n_{{l_b}'' \sigma'}{\rangle} + V_{\rm edge} ,
\end{eqnarray}
where $ {\langle} n_{l_b \sigma} {\rangle} = \sum_{k_a}{\langle}a^\dag_{k_a l_b \sigma}a_{k_a l_b \sigma} {\rangle}$ is the mean density of electrons with spin $\sigma$ at site $l_b$ averaged for the Fermi distribution. 
Diagonalization of $H_{\rm MF}$ leads to the energy eigenvalue $E_{\nu \sigma}(k_a)$ and the eigenfunction $d_{l_b \nu \sigma}(k_a)$, which results in a formation of \ohkimodmod{the} energy bands $\nu = 1, 2, {\cdots}, 4N_b-2$ $(1, 2, {\cdots}, 4N_b-1, 4N_b)$ for the \ohkimodmod{(AA$'$-AA$'$) symmetric-edge ((AA$'$-BC) asymmetric-edge)} pattern. 
Recalling the orthogonality $\sum_{k_a \nu \sigma}d_{l_b \nu \sigma}(k_a)d^*_{l_b \nu \sigma}(k_a) = \delta_{l_b {l_b}'}$, the Hamiltonian becomes 
\begin{eqnarray}
H_{\rm MF}= \sum_{k_a \nu \sigma} E_{\nu \sigma}(k_a) c^\dag_{k_a \nu \sigma}c_{k_a \nu \sigma}+{\rm const.} ,
\\
E_{\nu \sigma}(k_a)d_{l_b \nu \sigma}(k_a) = \sum_{{l_b}'} \tilde{\varepsilon}_{l_b {l_b}' \sigma}(k_a)d_{{l_b}' \nu \sigma}(k_a) ,
\end{eqnarray}
with $E_{1 \sigma}(k_a) < E_{2 \sigma}(k_a) <\cdots < E_{4N_b-2 \sigma}(k_a)$. 
The chemical potential $\mu$ is determined from the quarter filling condition $\frac{1}{4}\sum_{l_b \sigma}{\langle}n_{l_b \sigma}{\rangle} =\frac{3}{2}$. Note that we set $\hbar = k_{\rm B} = 1$. 
Using Eqs. (A6) and (A7), we will evaluate the electronic properties at finite temperature, in particular \ohkimodmod{the optical and dc} \ohkimodmod{conductivities}.

The optical conductivity along the \ohkimodmod{$a$} axis is calculated by the Nakano-Kubo formula which is formulated in terms of a linear response theory, given by
\begin{eqnarray}
\sigma^a(\omega)&=&\frac{1}{i\omega}\left[ Q^R(\omega) - Q^R(0) \right] ,\\
Q^R(\omega)
&=&-\frac{e^2}{\Omega}\sum_{k_a \nu \nu' \sigma}\left|{\rm v}^a_{\nu \nu' \sigma}(k_a) \right|^2 \nonumber\\
&&\times \frac{f(E_{\nu \sigma}(k_a))-f(E_{\nu' \sigma}(k_a))}{E_{\nu \sigma}(k_a)-E_{\nu' \sigma}(k_a)+\hbar \omega+i0^+} ,
\end{eqnarray}
\ohkimodmod{where} $\Omega = N_a \times N_b$ \ohkimodmod{is the 2D system size and $f(E_{\nu \sigma}(k_a))$ is} the Fermi-Dirac distribution function.

The longitudinal dc conductivity~\cite{Streda, Shon, Proskurin, Ruegg2008, Omori2017} along the \ohkimodmod{$a$} axis is given by 
\begin{eqnarray}
{\hspace{-0.9cm}}\sigma^a &=& \int d\omega \left( -\frac{d f}{d\omega} \right)\Phi(\omega) ,\\
{\hspace{-0.9cm}}\Phi(\omega) {\hspace{-0.1cm}}&=&{\hspace{-0.1cm}} \frac{2e^2}{\pi \Omega}\sum_{k_a}{\rm Tr}\left[ {\rm v}^a {\rm Im}{\hat G}^R(\omega, k_a){\rm v}^a{\rm Im}{\hat G}^R(\omega, k_a) \right] ,
\end{eqnarray}
where \ohkimodmod{we introduced} velocity variables 
\begin{eqnarray}
{\hspace{-0.7cm}}{\rm v}^a_{\nu \nu' \sigma}(k_a)&=&\sum_{l_b l_b'}d^*_{l_b \nu \sigma}(k_a)v^a_{l_b {l_b}' \sigma}(k_a)d_{{l_b}' \nu' \sigma}(k_a) ,\\
{\hspace{-0.7cm}}{\it v}^a_{l_b {l_b}' \sigma}(k_a)&=&\frac{\partial}{\partial k_a}\tilde{\epsilon}_{l_b {l_b}' \sigma}(k_a) .
\end{eqnarray}
\ohkimodmod{The retarded Green's function $\hat{G}^R$ is expressed} in the T-matrix approximation as 
\begin{eqnarray}
{\hspace{-0.7cm}}G^{R}_{l_b {l_b}' \sigma}(\omega, k_a) = \sum_{\nu}\frac{d_{l_b \nu \sigma}(k_a)d^*_{{l_b}' \nu \sigma}(k_a)}{\hbar \omega - E_{\nu \sigma}(k_a)-\Sigma^R_{\nu \sigma}(\omega, k_a)} 
\end{eqnarray}
with the retarded self-energy
\begin{eqnarray}
{\hspace{-0.7cm}}\Sigma^R_{\nu \sigma}(\omega, k_a) = \sum_{l_b}\frac{n_{\rm imp} V_{\rm imp}\left|d_{l_b \nu \sigma}(k_a) \right|^2}{1-\frac{V_{\rm imp}}{N_a}\sum_{{k_a}'}G^{0R}_{l_b l_b \sigma}(\omega, {k_a}')} 
\end{eqnarray}
and the single-particle retarded \ohkimodmod{Green's} function 
\begin{eqnarray}
{\hspace{-0.7cm}}G^{0R}_{l_b {l_b}' \sigma}(\omega, k_a) = \sum_{\nu}\frac{d_{l_b \nu \sigma}(k_a)d^*_{{l_b}' \nu \sigma}(k_a)}{\hbar \omega - E_{\nu \sigma}(k_a)+i0^+} .
\end{eqnarray}
\ohkimodmod{Here, we defined} the impurity potential $V_{\rm imp}$ 
\begin{eqnarray}
{\hspace{-0.7cm}}H' &=& \frac{V_{\rm imp}}{N_a}\sum_{k_a q_a l_b \sigma}\rho_{\rm imp}(q_a)a^\dag_{k_a+q_a l_b \sigma}a_{k_a l_b \sigma} ,\\
{\hspace{-0.7cm}}\rho_{\rm imp}(q_a) &=& \sum_{i_a}e^{-iq_a r_{i_a}} ,
\end{eqnarray}
for the number of impurity centers $N_{\rm imp}$ and the density of impurities $n_{\rm imp} = N_{\rm imp} / N_a$ in the \ohkimodmod{$a$} direction. 
The damping constant $\gamma_{\nu \sigma}(\omega, k_a)$ is given by the imaginary part of $\Sigma^R$ and is calculated as 
\begin{eqnarray}
{\hspace{-0.8cm}} \gamma_{\nu \sigma}(\omega, k_a) &=& -{\rm Im}\Sigma^R_{\nu \sigma}(\omega, k_a) \nonumber\\
&=& \sum_{l_b}\frac{\left| d_{l_b \nu \sigma}(k_a) \right|^2\left[ \pi n_{\rm imp} V_{\rm imp}^2 \mathcal{N}_{l_b \sigma}(\omega)\right]}{1+\left[ \pi V_{\rm imp}\mathcal{N}_{l_b \sigma}(\omega) \right]^2} ,
\end{eqnarray}
where \ohkimodmod{$\mathcal{N}_{l_b \sigma}(\omega)$ is a site-resolved} spectral density \ohkimodmod{given by}
\begin{eqnarray}
{\hspace{-0.5cm}} \mathcal{N}_{l_b \sigma}(\omega) &=& \frac{1}{N_a}\sum_{k_a}\left( -\frac{1}{\pi} {\rm Im}G^{0R}_{l_b l_b \sigma}(\omega, k_a) \right) \nonumber\\
&=& \frac{1}{N_a}\sum_{k_a \nu}\delta(\hbar \omega - E_{\nu \sigma}(k_a))\left| d_{l_b \nu \sigma}(k_a) \right|^2 .
\end{eqnarray}
Recalling \ohkimodmod{that} $\tau = 1/2\gamma$ defines the relaxation time, one is able to simplify $\Phi(\omega)$ in Eq. (A11) \ohkimodmod{as follows:} 
\begin{equation}
\Phi(\omega)=\frac{4e^2}{\Omega}\sum_{k_a \nu}\left| {\rm v}^a_{\nu}(k_a) \right|^2 \tau_\nu (\omega, k_a)\delta(\hbar \omega - E_\nu (k_a)) .
\end{equation}
\ohkimodmod{To} see the spatial distribution of conductivities along the $b$ direction, a spatially-resolved conductivity \ohkimodmod{is defined by} 
\begin{eqnarray}
\sigma^a_{i_b} = \int d\omega\left( -\frac{df}{d\omega} \right)\Phi_{i_b}(\omega) ,
\end{eqnarray}
with a \ohkimodmod{spatially-resolved} distribution function
\begin{eqnarray}
\Phi_{i_b}(\omega) &=& \sum_{\alpha {l_b}'}\Phi_{l_b {l_b}'}(\omega) ,\\
\Phi_{l_b {l_b}'}(\omega) &=& \frac{4e^2}{N_a}\sum_{k_a \nu}{\rm v}^a_\nu (k_a)\left[ d^*_{l_b \nu}(k_a)v^a_{l_b {l_b}'}(k_a)d_{{l_b}' \nu}(k_a) \right]\nonumber\\
&&\times \hspace{0.2cm} \tau_\nu (\omega, k_a)\delta(\hbar \omega - E_\nu (k_a)) .
\end{eqnarray}
The dc conductivity $\sigma^a$ (or \ohkimodmod{the} dc resistivity $\rho^a = 1/\sigma^a$ ) \ohkimodmod{is} derived from a summation of $\sigma^a_{i_b}$ over \ohkimodmod{all possible} $i_b$ \ohkimodmod{in the $b$ direction}.



\begin{thebibliography}{}
\bibitem{Kajita1992}
K. Kajita, T. Ojiro, H. Fujii, Y. Nishio, H. Kobayashi, A. Kobayashi, and R. Kato, J.\ Phys.\ Soc.\ Jpn.\ \textbf{61}, 23 (1992). 
\bibitem{Tajima2000}
N. Tajima, M. Tamura, Y. Nishio, K. Kajita, and Y. Iye,
J.\ Phys.\ Soc.\ Jpn.\ \textbf{69}, 543-551 (2000).
\bibitem{Kobayashi2004}
A. Kobayashi, S. Katayama, K. Noguchi, and Y. Suzumura, J.\ Phys.\ Soc.\ Jpn.\ \textbf{73}, 3135 (2004)
\bibitem{Katayama2006}
S. Katayama, A. Kobayashi, and Y. Suzumura,
J.\ Phys.\ Soc.\ Jpn.\ \textbf{75}, 054705 (2006).
\bibitem{Kobayashi2007}
A. Kobayashi, S. Katayama, Y. Suzumura, and H. Fukuyama,
J.\ Phys.\ Soc.\ Jpn.\ \textbf{76}, 034711 (2007).
\bibitem{Goerbig2008}
M. O. Goerbig, J.-N. Fuchs, G. Montambaux, and F. Pi$\acute{{\rm e}}$chon,
Phys.\ Rev.\ B \textbf{78}, 045415 (2008).
\bibitem{Kajita2014}
K. Kajita, Y. Nishio, N. Tajima, Y. Suzumura, and A. Kobayashi, J.\ Phys.\ Soc.\ Jpn.\ \textbf{83}, 072002 (2014).
\bibitem{KinoFukuyama}
H. Kino and H. Fukuyama,
J.\ Phys.\ Soc.\ Jpn.\ \textbf{64}, 4523 (1995).
\bibitem{Seo2000}
H. Seo,
J.\ Phys.\ Soc.\ Jpn.\ \textbf{69}, 805 (2000).
\bibitem{TakahashiStripe}
T. Takahashi, Synth.\ Met.\ \textbf{133-134}, 26 (2003).
\bibitem{Kakiuchi2007}
T. Kakiuchi,  Y. Wakabayashi,  H. Sawa,  T. Takahashi, and  T. Nakamura, J. Phys. Soc. Jpn. 76, 113702 (2007).
\bibitem{Bender1984}
K. Bender ,I. Hennig ,D. Schweitzer ,K. Dietz ,H. Endres and H. J. Keller, Molecular Crystals and Liquid Crystals. \textbf{108}, 359-371 (1984).
\bibitem{Tanaka2016}
Y. Tanaka and M. Ogata, JPSJ 85, 104706 (2016).
\bibitem{Ishikawa2016}
{K. Ishikawa, M. Hirata, D. Liu, K. Miyagawa, M. Tamura, and K. Kanoda, \ Phys.\ Rev.\ B \textbf{94}, 085154 (2016).}
\bibitem{Matsuno2016}
G. Matsuno, Y. Omori, T. Eguchi, and A. Kobayashi: J.\ Phys.\ Soc.\ Jpn.\ \textbf{85} 094710 (2016).
\bibitem{Omori2017}
Y. Omori, G. Matsuno, A. Kobayashi, J.\ Phys.\ Soc.\ Jpn.\ \textbf{86}, 074708 (2017).
\bibitem{Ohki2018JPSJ}
D. Ohki, G. Matsuno, Y. Omori, and A. Kobayashi, J. Phys. Soc. Jpn. \textbf{87}, 054703 (2018).
\bibitem{Ohki2018Crystals}
D. Ohki, G. Matsuno, Y. Omori, and A. Kobayashi ,Crystals \textbf{2018}, 8 (3), 137.
\bibitem{Kobayashi2008IOP}
{A. Kobayashi, S. Komaba, S. Katayama, and Y. Suzumura, J. Phys.: Conf. Ser. \textbf{132} 012002 (2008). }
\bibitem{Dietl}
{P. Dietl, F. Pi\'{e}chon, and G. Montambaux, Phys. Rev. Lett. \textbf{100}, 236405 (2008). }
\bibitem{Montambaux2009PRB}
{G. Montambaux, F. Pi\'{e}chon, J. -N. Fuchs, and M. O. Goerbig, Phys. Rev. B \textbf{80}, 153412 (2009).}
\bibitem{Montambaux2009EPJB}
{G. Montambaux, F. Piechon, J.-N. Fuchs, and M. O. Goerbig, Eur. Phys. J. B \textbf{72}, 509 (2009).}
\bibitem{Kobayashi2011}
{A. Kobayashi, Y. Suzumura, F. Piechon, and G. Montambaux, Phys. Rev. B \textbf{84}, 075450 (2011). }
\bibitem{Hirata2016}
M. Hirata, K. Ishikawa, K. Miyagawa, M. Tamura, C. Berthier, D. Basko, A. Kobayashi, G. Matsuno and K. Kanoda, Nat. Commun. \textbf{7}, 12666 (2016). 
\bibitem{Hirata2017}
M. Hirata, K. Ishikawa, G. Matsuno, A. Kobayashi, K. Miyagawa, M. Tamura, C. Berthier, K. Kanoda,  Science \textbf{358}, 1403 (2017).
\bibitem{Matsuno2017}
G. Matsuno and A. Kobayashi, J.\ Phys.\ Soc.\ Jpn.\ \textbf{86} 014705 (2017).
\bibitem{Matsuno2018}
{G. Matsuno and A. Kobayashi, J. Phys. Soc. Jpn. \textbf{87}, 054706 (2018).}
\bibitem{Liu2016}
D. Liu, K. Ishikawa, R. Takehara, K. Miyagawa, M. Tamura,
and K. Kanoda: Phys. Rev. Lett. \textbf{116} 226401 (2016).
\bibitem{Beyer2016}
R. Beyer, A. Dengl, T. Peterseim, S. Wackerow, T. Ivek, A.V. Pronin, D. Schweitzer, and M. Dressel, Phys.\ Rev.\ B \textbf{93}, 195116 (2016).
\bibitem{Yamamoto2008}
{Kaoru Yamamoto, Shinichiro Iwai, Sergiy Boyko, Akimitsu Kashiwazaki, Fukiko Hiramatsu, Chie Okabe, Nobuyuki Nishi, and Kyuya Yakushi, J. Phys. Soc. Jpn. \textbf{77}, 074709 (2008).}
\bibitem{Yamamoto2010}
{Kaoru Yamamoto, Aneta Aniela Kowalska, and Kyuya Yakushi, Appl. \ Phys.\ Lett. \textbf{96}, 122901 (2010).}
\bibitem{Lunkenheimer}
{P. Lunkenheimer, B. Hartmann, M. Lang, J. Muller, D. Schweitzer, S. Krohns, and A. Loidl,  \ Phys.\ Rev.\ B \textbf{91}, 245132 (2015). }
\bibitem{Hasegawa2011}
Y. Hasegawa and K. Kishigi, J.\ Phys.\ Soc.\ Jpn.\ \textbf{80}, 054707 (2011).
\bibitem{Omori2014}
Y. Omori, G. Matsuno, and A. Kobayashi, JPS Conf.\ Proc.\  \textbf{1}, 012119 (2014).
\bibitem{Kobayashi2009IOP}
A. Kobayashi, S. Katayama, and Y. Suzumura, Sci. Technol. Adv. Mater. \textbf{10}, 024309 (2009).
\bibitem{Tajima2009}
N. Tajima and K. Kajita, Sci. Technol. Adv. Mater. 10, 024308 (2009).
\bibitem{Streda}
P. St$\breve{\rm r}$eda and L. Smr$\breve{\rm c}$ka, Phys. stat. sol. (b) \textbf{70}, 537 (1975).
\bibitem{Shon}
N. H. Shon and T. Ando, J. Phys. Soc. Jpn. \textbf{67} 2421 (1998).
\bibitem{Proskurin}
I. Proskurin, M. Ogata, and Y. Suzumura, Phys. Rev. B \textbf{91} 195413 (2015).
\bibitem{Ruegg2008}
{A. R$\rm {\ddot{u}}$egg, S. Pilgram, and M. Sigrist, Phys. Rev. B \textbf{77} 245118 (2008).}
\bibitem{Kondo2009}
R. Kondo, S. Kagoshima, N. Tajima, and R. Kato, J. Phys. Soc. Jpn. \textbf{78}, 114714 (2009).
\bibitem{Alemany}
P. Alemany, J. Pouget, and E. Canadell, Phys. Rev. B \textbf{85}, 195118 (2012).
\bibitem{Kino}
H. Kino and  T. Miyazaki, J.\ Phys.\ Soc.\ Jpn.\ \textbf{75}, 034704 (2006). 
\bibitem{Ivek2017}
T. Ivek, M. \u{C}ulo, M. Kuve\u{z}di\'{c}, E. Tuti\u{s}, M. Basleti\'{c}, B. Mihaljevi\'{c}, E. Tafra, S. Tomi\'{c}, A. L\"{o}hle, M. Dressel, D. Schweitzer, and B. Korin-Hamzi\'{c}, Phys. Rev. B \textbf{96}, 075141 (2017). 



\end{thebibliography}


\end{document}